\documentclass[%
 aip,
 amsmath,amssymb,
 reprint,%
]{revtex4-1}

\usepackage{graphicx}
\usepackage{dcolumn}
\usepackage{bm}

\usepackage[utf8]{inputenc}
\usepackage[T1]{fontenc}
\usepackage{mathptmx}
\usepackage{etoolbox}
\usepackage{color}

\makeatletter
\def\@email#1#2{%
 \endgroup
 \patchcmd{\titleblock@produce}
  {\frontmatter@RRAPformat}
  {\frontmatter@RRAPformat{\produce@RRAP{*#1\href{mailto:#2}{#2}}}\frontmatter@RRAPformat}
  {}{}
}%
\makeatother
\begin{document}

\preprint{AIP/123-QED}

\title[DFT+$\mu$ for muon site determination]{DFT+$\mu$: Density Functional Theory for Muon Site Determination}
\author{S. J. Blundell}
\affiliation{Department of Physics, Clarendon Laboratory, Oxford University,
Parks Road, Oxford OX1 3PU, United Kingdom}
\author{T. Lancaster}%
\affiliation{ 
Department of Physics, Centre for Materials Physics,
Durham University, Durham, DH1 3LE, United Kingdom
}%
\email{stephen.blundell@physics.ox.ac.uk and tom.lancaster@durham.ac.uk}

\date{\today}

\begin{abstract}
The technique of muon spin rotation ($\mu$SR) has emerged in the last few decades as one of the most powerful methods of obtaining local magnetic information.  To make the technique fully quantitative, it is necessary to have an accurate estimate of where inside the crystal structure the muon implants.  This can be provided by density functional theory calculations using an approach that is termed DFT+$\mu$, density functional theory with the implanted muon included.  This article reviews this approach, describes some recent successes in particular $\mu$SR experiments, and suggests some avenues for future exploration.
\end{abstract}

\maketitle

\section{Introduction}\label{sec:intro}
Very often in condensed matter physics it is necessary to have very detailed information about the magnetic properties of materials measured at a local level.  Standard magnetic characterisation can be provided by measurements of magnetic susceptibility $\chi$; this quantifies the magnetic response of a sample, revealed by its magnetic moment $m$ induced by an applied field $H$, and ideally extracted in the limit of
$H\to 0$, so that $\chi=\lim_{H\to 0}M/H$, where $M$ is the magnetization $M=m/V$ and $V$ is the sample's volume.  This has two obvious problems: (1) the limit of $H\to 0$ is hard to achieve since the magnitude of the measurement signal is often proportional to $H$, so measurements have to be performed in non-zero, and sometimes substantial, applied field; (2) the measurement averages over the entire sample volume, since $m=\int_V M\,{\rm d}V$, and so there is no way to distinguish between a sample that is uniformly ordered, but with a small $M$, and one that is not ordered, but has a small component which is ordered with a large $M$.  In such cases, what is needed is 
a highly-sensitive {\it local} magnetic probe, ideally one operating at the atomic level and giving single spin detection.

Spin-polarized muons provide exactly this kind of sensitive local magnetic probe.\cite{Cox1987,Reotier1997,blundell1999,blundell2022} In a muon-spin rotation ($\mu$SR) experiment, a beam of spin-polarized muons is implanted in the sample to be studied. When each muon decays, a positron and two neutrinos are emitted. A property of this decay is that the positron is not emitted isotropically, but preferentially along the direction of the muon spin at the moment of decay.\cite{Garwin1957} Detecting the direction along which the positron is emitted allows one to infer which direction the muon spin was pointing at the moment of its decay. Each muon lives for a different amount of time (according to the radioactive decay law) and so each muon decay gives rise to a positron whose detection at time $t$ at a particular angle of emission contributes to a data point. Making many such positron detections, taken over many million muon decays, allows one to build up a histogram $A(t)$, yielding the average muon spin polarization $P(t)$ of the large ensemble of muons as a function of time, allowing both static and dynamic local fields to be probed. 

The technique has a number of key advantages.  First of all, muon
beams have 100\% spin polarization (owing to parity violation in the weak interaction), in contrast to the very weak thermal polarizations obtained in nuclei in nuclear magnetic resonance (NMR) experiments.
Secondly, muons have a larger gyromagnetic ratio than any nucleus in NMR, resulting in high sensitivity.
It is very helpful that
there is no restriction or necessity for specific nuclear isotopes, with muons usable for measurements
in any material (in contrast to both NMR and neutron scattering).
The ability to use an extended field range for measurements, from zero field (ZF) up to 10 T, and a wide range of temperatures (using dilution refrigerators, helium cryostats or furnaces), means that $\mu$SR experiments are compatible with a very wide range of sample environment.
Finally, the technique allows the user to study the time dependence of local magnetic fields and the technique gives
access to dynamical information with a very wide range of correlation times ($10^{-10}$--$10^{-5}$~s) that fit neatly between a.c.\ susceptibility\cite{Topping2019} on one hand and neutron scattering\cite{Boothroyd2021} on the other (the longest times are limited by the muon lifetime and the shortest times depend on the local field distribution \cite{blundell2022}).  Consequently, it has been widely used to study problems in magnetism,\cite{Reotier1997} superconductivity,\cite{Sonier2000} organic conductors,\cite{Blundell2004} and many other systems (for an up-to-date review of applications of $\mu$SR to a very wide range of topics, see Ref.~\onlinecite{blundell2022}), and has become a frequently deployed method alongside other experimental techniques, such as magnetic neutron diffraction.\cite{Boothroyd2021}

Despite many successes of the technique, there is a perceived drawback of the technique that arises from the lack of knowledge of the muon's stopping site in materials, raising the question of which magnetic fields are actually being probed. 
Consider
  the dipolar field $\boldsymbol{B}_{\rm dip}$ measured by the muon, resulting from the ordered magnetic moments in the sample.  The $\alpha$-component of the dipolar
  ($\alpha=x$, $y$ or $z$)
  is given by
  \begin{equation}
    B_{\rm dip}^\alpha  = \sum_j \sum_\beta D^{\alpha\beta}_j m^\beta_j, 
    \label{eq:dipole1}
  \end{equation}
  where the first sum is taken over all the magnetic moments
  $\boldsymbol{m}_j$ within
  the Lorentz sphere (i.e.\ out to a large enough radius from the muon) 
  and the dipolar tensor for the $j$th moment
  at position $\boldsymbol{r}_j$ is given by
  \begin{equation}
    D^{\alpha\beta}_j = {\mu_0 \over 4 \pi R_j^3} \left( 
    { 3 R_j^\alpha R_j^\beta \over R_j^2 } - \delta^{\alpha\beta} \right),
    \label{eq:dipole2}
  \end{equation}
  where $\boldsymbol{R}_j = \boldsymbol{r}_j - \boldsymbol{r}_\mu$ 
  and $\boldsymbol{r}_\mu$ is the muon position.  These expressions
  are often used to model ordered magnetic arrangements and this can
  be a very good test of the validity of a muon site, but notice that the expressions depend on the coordinates $\boldsymbol{R}_j$ which are measured with respect to the (unknown) muon position 
  $\boldsymbol{r}_{\mu}$.
There are other contributions to the field at the muon site in addition to $\boldsymbol{B}_{\rm dip}$ (such as the hyperfine field and, in the case of ferromagnets, the Lorentz and demagnetization fields\cite{blundell2022}), but $\boldsymbol{B}_{\rm dip}$ is the most significant and illustrates the point that the field at the muon site is a function of the unknown quantity
$\boldsymbol{r}_{\mu}$.

Even more serious is the unknown effect that the positively charged muon has on its local environment, raising the question of whether intrinsic behaviour is being measured, or whether instead the experimental results are caused by the presence of the implanted muon itself. One can suspect that, if the final muon stopping state involves relatively little contact hyperfine coupling, the stopped muon will resemble something like a bare particle playing the role of an interstitial defect, albeit ‘dressed’ by interactions with the electronic system. This might involve, for example, the muon acquiring a screening cloud of electronic charge in a metal, or acquiring a strain field in an insulator as it deforms atoms in its vicinity. Knowledge of the muon’s stopping site in such cases would allow us insight into the local fields and interactions at a known position. If the muon instead forms a bound paramagnetic state such as muonium, or induces a large spin density at its position, knowledge of the muon site would therefore allow us to identify the electronic state that gives rise to the measured hyperfine coupling constants, and hence direct
insight into the local electronic structure.
To address this drawback of $\mu$SR, a density functional theory (DFT) ab-initio approach has been developed to determine
the muon site.\cite{Moller2013,Bernardini2013,Moller2013b,Bonfa2016} We call this approach DFT$+\mu$, density functional theory with an implanted muon, and this review will describe the method and demonstrate some of its recent successes.

\section{The site problem}
The  location of the muon site is determined by all the electrostatic interactions in the system, including the distortions produced in the material by the presence of the implanted muon, and interactions between all the particles in the system.  This is a complicated problem, but is pretty well described by non-relativistic quantum mechanics.
The system comprising the stopped muon and its host material can therefore be described by the Schr\"{o}dinger equation, and so a computation of the wavefunction of the system should, in principle, be possible. The system is however rather large, comprising the muon and the large number of electrons and nuclei in the sample.  This is a many body problem and therefore extremely complex.  Even if we restrict ourselves to focussing only on the electronic part of the wave function (which makes sense, as it is the behaviour of the more mobile electrons that ultimately determines the muon site) the task is formidable.\cite{Kohn1999}  In Section~\ref{sec:dftmu} we will describe how this problem can be treated with electronic structure calculations, but in the remainder of this section we will describe some methods that have been used previously.

\subsection{The intuitive approach}
Even without the use of sophisticated electronic structure calculations, it is possible to think about where a muon is likely to sit using chemical intuition.  Most $\mu$SR experiments use positive muons ($\mu^+$) and since the muon behaves like a light isotope of hydrogen, the $\mu^+$ is likely to implant at a site that would be favoured by the H$^+$ cation.  Therefore, one can immediately deduce some
ad hoc rules of thumb for guessing muon sites. For a start, one can conclude that positive muons favour sites close to negatively charged anions.  A good example is provided by the large number of magnetic oxides that have been widely studied in condensed matter physics. In these it is usually found that the muon sits around 1\,\AA\ from an O$^{2-}$ ion (a result which was concluded during studies of the cuprate superconductors\cite{Brewer1991} which all consist of a collection of cations and oxygen anions).  The resulting state behaves like a hydroxyl OH group, which has a bond length close to 1\,\AA.\cite{Demaison2007}
In fluorides, it is known that muons form F$-\mu-$F bonds\cite{Brewer1986} in which the muon sits between the two fluoride ions, forming an analogue of the bifluoride ion (HF$_2^-$).  We will return to this important example later.   
In elemental metals, the muon would be expected to occupy a highly symmetrical site, such as an interstitial site inside the face-centred cubic lattice of copper for example.\cite{Luke1991}
These empirical rules of thumb are often useful, and lead to the identification of highly plausible muon sites, albeit in a limited range of materials. 

\subsection{Bayesian methods}
One fruitful approach to solving (or at least avoiding) the muon site problem is to deny that knowledge of the muon site is necessary.  If the goal is to measure a magnetic moment of a magnetic ion in a structure, is the muon site really needed and can one include our quantified ignorance of it in the calculation?  This approach leads to a 
Bayesian method\cite{Blundell2012} and is based on the observation that evaluating the dipolar field, given the muon site and the magnetic moment, is an easy problem (just use eqns~\ref{eq:dipole1} and \ref{eq:dipole2}), but the reverse problem is hard.  The inversion between the two calculations is accomplished using Bayes' theorem which states that
$P({\sf A} \vert {\sf B}) =  P({\sf A}) P({\sf B} \vert {\sf A}) / P({\sf B})$.
Here $P({\sf A})$ is called the {\sl prior probability}, since it is the
probability of ${\sf A}$ occurring without any knowledge as to the outcome
of ${\sf B}$.  The quantity which you derive is
$P({\sf A} \vert {\sf B})$, the {\sl posterior probability}.
For the muon problem, we write Bayes' theorem as
\begin{equation}
P(\mu \vert \nu ) = { P(\mu) P( \nu \vert \mu )  \over \int P ( \nu
  \vert \mu') P(\mu') \,{\rm d}\mu' },
\label{eq:bayes}
\end{equation}
where $\nu$ is a muon precession frequency and $\mu$ is the magnetic moment of an ion in the crystal.  Equation~\ref{eq:bayes} gives the probability of a magnetic moment, given the observed muon precession frequency (the experimental problem, which is hard), in terms of the probability distribution of the muon precession frequency, given the size of the magnetic moment (the computational problem, which is easy).  These are probabilities, since we do not know the muon site {\it a priori} and so the muon site is described by some probability distribution within the unit cell (and for the case of total ignorance of the muon site, the probability distribution can be uniform throughout the unit cell).  Thus we can compute the distribution of dipolar fields within the unit cell under some assumed distribution of muon sites (reflecting the level of our ignorance), which we can express as a probability density function (pdf)
$f(\nu/\mu)$, evaluated as a function of precession frequency $\nu$
divided by magnetic moment $\mu$ (since the precession frequency
scales with the magnetic moment).  This function $f(\nu/\mu)$
  allows us to evaluate $P(\nu\vert\mu)$ of eqn~(\ref{eq:bayes}).  Thus
$P(\nu\vert\mu)={1\over \mu}f(\nu/\mu)$, and since
$f(\nu/\mu)$ is normalized so that $\int f(\nu/\mu)\,{\rm
  d}(\nu/\mu)=1$, 
the factor of ${1\over \mu}$ is needed
so that
$\int P(\nu\vert\mu)\,{\rm d}\nu=1$

Since $\nu$ is obtained from a real experiment, what we would like to know is
$g(\mu\vert\nu)$, the pdf of $\mu$ {\it given} the
observed $\nu$.  This can be obtained from our calculated $f(\nu/\mu)$
using Bayes' theorem in the form of Eq.~(\ref{eq:bayes}), which yields
\begin{equation}
g(\mu\vert\nu) = { { 1 \over \mu} f(\nu/\mu) \over \int_0^{\mu_{\rm max}} 
{1 \over \mu'} f(\nu/\mu')\,{\rm d}\mu' }, 
\label{eq:final}
\end{equation}
where we have assumed a prior probability [$P(\mu)$] for the magnetic
moment that is uniform between zero and $\mu_{\rm max}$, and so
$P(\mu)$ is replaced by the uniform probability density $1/\mu_{\rm
  max}$ [which cancels on the top and bottom of Eq.~(\ref{eq:final})].  We
choose $\mu_{\rm max}$ to take a large value, although it is found\cite{Blundell2012}
that results are insensitive to the precise value of $\mu_{\rm
  max}$.  
When multiple frequencies $\nu_{i}$ are present in the
spectra, it is necessary to multiply their probabilities of
observation in order to obtain the chance of their simultaneous
observation, so we evaluate $g(\mu\vert\{\nu_i\}) \propto \prod_{i}
\int_{\nu_i-\Delta\nu_i}^{\nu_i+\Delta\nu_i}f(\nu_i/\mu)\,{\rm
  d}\nu_i$, where $\Delta\nu_i$ is the error on the fitted frequency.
These results have been applied successfully in a number of studies.\cite{Blundell2012,Steele2011,Disseler2014,Prando2020}

\subsection{Knight shift measurements}
An experimental method that has been used to identify or verify a muon site involves applying a transverse 
magnetic field $\boldsymbol{B}_{\rm ext}$ to the sample and measuring the Knight shift.
This occurs because the field at  the muon site
$\boldsymbol{B}_\mu$ may not be exactly the same as the applied field
$\boldsymbol{B}_{\rm ext}$, due to contributions from the
dipolar field, the hyperfine contact interaction, the Lorentz field,
and the demagnetization field.  In a transverse field experiment, one
is often interested in the difference between the two,
$\boldsymbol{B}_\mu-\boldsymbol{B}_{\rm ext}$, but measured along
$\boldsymbol{B}_{\rm ext}$.  This is because the component of
$\boldsymbol{B}_\mu-\boldsymbol{B}_{\rm ext}$ measured perpendicular to
$\boldsymbol{B}_{\rm ext}$ makes very little difference to the
precession frequency. The Knight shift
$K$ is then defined by
\begin{equation}
K = { (\boldsymbol{B}_\mu-\boldsymbol{B}_{\rm ext})\cdot
  \boldsymbol{B}_{\rm ext}
  \over B_{\rm ext}^2 },
\end{equation}
and hence $B_\mu \approx (1+K) B_{\rm ext}$.
The dipolar field from ordered moments contributes to $K$, but
 even in
  the paramagnetic state 
  the moments can become partially polarized in an applied field and take the value
  \begin{equation}
    \boldsymbol{m}_j = { \underline{\boldsymbol{\chi}}
      \boldsymbol{B}_{\rm ext} {\cal V}_{\rm c} \over \mu_0 },
  \end{equation}
  where ${\cal V}_{\rm c}$ is the volume per magnetic ion and
  $\underline{\boldsymbol{\chi}}$ is the magnetic susceptibility
  tensor.  Hence
\begin{equation}
\boldsymbol{B}_{\rm dip} = 
\underline{\boldsymbol{{\cal D}}}
\underline{\boldsymbol{\chi}}
\boldsymbol{B}_{\rm ext},
\end{equation}
where
  $\underline{\boldsymbol{{\cal D}}} = {{\cal V}_{\rm c} \over \mu_0}\sum_j
  \underline{\boldsymbol{D}}_{j}$ is the total dipolar tensor.  Similarly,
the contact hyperfine interaction
  $\boldsymbol{B}_{\rm hf}$ is given by
\begin{equation}
\boldsymbol{B}_{\rm hf} = 
\underline{\boldsymbol{{\cal A}}}
\underline{\boldsymbol{\chi}}
\boldsymbol{B}_{\rm ext},
\end{equation}
where
  $\underline{\boldsymbol{{\cal A}}}$ is the analogous hyperfine tensor.
  The contact coupling is usually independent of the field direction
  and so this can be written as a scalar in the majority of cases.
This provides us with all we need to model the effect of rotating a
  single crystal in a constant magnetic field and predicting the
  Knight shift for a particular muon site,
  the anisotropy of the dipolar coupling resulting in
  angle dependence, while the contact interaction gives an
  angle-independent contribution.
  
  Another experimental test of the validity of a candidate muon site can
be obtained using the contribution of nearby nuclear dipoles to the
decay rate of a transverse-field precession measurement.  The damping of the
precession signal arises from contributions from nearby nuclear
dipoles, which are not ordered, but sometimes add and sometimes
subtract from the applied field.  In a transverse field measurement,
the broadening is given at short times by a Gaussian relaxation function
$\exp(-\sigma^2 t^2 / 2)$, where the parameter $\sigma^2$ (also known
as the second moment $M_2$ in NMR) is given by\cite{VanVleck1948,Abragam1961,Slichter1990,blundell2022}
\begin{equation}
\sigma^2 = \frac{1}{3} \left( {\mu_0 \over 4\pi} \right)^2 \hbar^2
\gamma_\mu^2
\sum_i \gamma_i^2 I(I+1) { (1-3\cos^2\theta_i)^2 \over r_i^6 }.
\end{equation}
This depends on the spin $I$ of the nearby nuclei, and their position
$r_i$ and gyromagnetic ratio $\gamma_i$.
This depolarization rate  is angle-dependent ($\theta_i$ is
the angle between the applied field and the vector between the muon
and the nucleus) and so this broadening can be measured
experimentally.
In the case of a zero-field measurement the relaxation takes on the
Kubo-Toyabe form with a value of $\Delta^2$ given by
\begin{equation}
\Delta^2 = \frac{1}{3} \left( {\mu_0 \over 4\pi} \right)^2 \hbar^2
\gamma_\mu^2
\sum_i \gamma_i^2 I(I+1) { (5-3\cos^2\theta_i) \over r_i^6 },
\end{equation}
which again can be checked by experiment.  Moreover, in some fortunate cases
(particularly fluorides) the zero-field signal from the nuclei
contains much more structure than a simple Kubo-Toyabe relaxation
(which is derived from an assumption of a Gaussian-distributed random
distribution of local field components).  In those cases, there is
much more information to go on to deduce the muon site and
understand the local environment.

\section{DFT+$\mu$}\label{sec:dftmu}
The methods described in the previous section can all be useful in guessing and experimentally checking the muon site.  However, electronic structure calculations are now proving to be extremely reliable as an ab initio technique for deducing muon sites, as well as for calculating the resulting distortion induced by the presence of the muon.
In this section, we first review briefly the ideas behind density functional theory (DFT) (for an excellent introductory review, see the book by Giustino\cite{Giustino2014}) and then describe how DFT can be implemented for muon site calculations.

\subsection{Density functional theory}
The many body wave function is a complex object to evaluate and store (making demands on memory size that grow exponentially with system size, resulting in one of the pioneers of the subject even doubting whether the many body wave function is a legitimate scientific concept for systems of more than a handful of atoms!\cite{Kohn1999}).  The approach taken in DFT\cite{Giustino2014,Kohn1999} is to dispense with the many body wave function and instead deal
with the functional $E[n(\boldsymbol{r})]$ whose output is the ground state energy of the $N$-electron system and whose input is the electron density function $n(\boldsymbol{r})$. 
The electron density is not a function of all of the coordinates of all of the electrons; it is simply a function of the three position coordinates $\boldsymbol{r}=(x,y,z)$; thus we treat the electron density as a fluid, and forget that the electrons are really individual quantum objects obeying antisymmetry conditions. 
The functional is therefore a quantity averaged over $3N-3$ of the degrees of freedom of the $N$-particle wavefunction. 
It seems surprising that we can make such a severe simplification and still compute the energy of a system, but that this is possible is the result of
the two  Hohenberg-Kohn (HK) theorems.\cite{HohenbergKohn1964}

The first HK theorem establishes a one-to-one correspondence
between an external potential $V(\boldsymbol{r})$ and the ground state
electronic density $n(\boldsymbol{r})$. It can be proved that the ground state
density $n(\boldsymbol{r})$ uniquely {\it determines} the
potential, and hence {\it all} of the
properties of the system.  So, for example, the
ground state energy $E$, can be written as a unique functional of the electron density $n(\boldsymbol{r})$ via an expression of the form
\begin{align}
E[n(\boldsymbol{r})]  = & T[n(\boldsymbol{r})] + E_{\mathrm{ee}}[n(\boldsymbol{r})] + E_{\mathrm{en}}[n(\boldsymbol{r})]\nonumber\\
                                  = & F[n(\boldsymbol{r})] + \int\mathrm{d}^{3}r\,n(\boldsymbol{r})V_{\mathrm{en}}(\boldsymbol{r}),
\label{eq:Enr}
\end{align}
where, in the final line of eqn~\ref{eq:Enr}, we have separated out the electron-nuclear part of the energy $E_{\mathrm{en}}$
and have written the rest as $F[n(\boldsymbol{r})]= T[n(\boldsymbol{r})] + E_{\mathrm{ee}}[n(\boldsymbol{r})]$, which is a functional encoding all of the purely electronic contributions to the energy.
If we have access to $E[n(\boldsymbol{r})]$ then we might be able to solve the problem of finding the electronic configuration corresponding to it by minimizing $E$ with respect to the electron density $n$. 
That is, we continually adjust the function $n(\boldsymbol{r})$ until we find a form that gives us the lowest-energy electron density $E$. 
Is this a safe strategy? The second HK theorem tells us that it is.

The second HK theorem states that $E[n(\boldsymbol{r})]$ gives the lowest
energy if, and only if, the input density $n(\boldsymbol{r})$ is the true ground state electron density. If we do not use the true density, then we obtain an upper bound on the ground state energy. 
Thus the idea is to minimize the functional $E[n(\boldsymbol{r})]$ with trial functions $n(\boldsymbol{r})$ and
if we find the actual minimum then we will have the true electron density. 
There is, however, a problem, in that  we do not actually know the exact form of 
the electron functional $F[n(\boldsymbol{r})]$. So, how do we find it?

A major simplification of the calculation of the electron density is made if we use the Kohn-Sham formulation,\cite{KohnSham1965}  the idea of which is to reformulate the problem by replacing the {\it interacting} system of many electrons by a {\it non-interacting} system of many electrons constrained to have the same electron density. 
Thus, the idea is to approximate the interacting kinetic energy $T[n(\boldsymbol{r})]$ of the $N$ interacting particles by the kinetic energy $T_{\mathrm{s}}[n(\boldsymbol{r})]$ of $N$ non-interacting particles with the same density $n(\boldsymbol{r})$.
We also extract from $F[n(\boldsymbol{r})]$ the Hartree component $U[n(\boldsymbol{r})]$. 
The Hartree component is simply the classical Coulomb interaction between regions of charge density, and can be easily written
in terms of $n(\boldsymbol{r})$. We then write 
\begin{equation}
F[n(\boldsymbol{r})] = T_{\mathrm{s}}[n(\boldsymbol{r})] + U[n(\boldsymbol{r})] +E_{\mathrm{xc}}.
\end{equation}
We have therefore bundled up our remaining ignorance into the functional 
 $E_{\mathrm{xc}}$, known as the exchange-correlation
energy.
To summarize, the HK theorems tell us we can find the ground state energy by minimizing a functional
$E[n(\boldsymbol{r})] = F[n(\boldsymbol{r})] + \int\mathrm{d}^{3}r\, V_{\mathrm{en}}(\boldsymbol{r})n(\boldsymbol{r})$ with respect to the density $n(\boldsymbol{r})$.
The KS formulation, with its replacement of the interacting system with an equivalent non-interacting one,
allows us to say that this minimization will be equivalent to a simpler procedure where we minimize the total energy of a non-interacting system subject to an effective potential $V_{\mathrm{s}}$, whose form is
\begin{equation}
V_{\mathrm{s}}(\boldsymbol{r}) = V_{\mathrm{en}}(\boldsymbol{r}) + V_{\mathrm{H}}(\boldsymbol{r}) + V_{\mathrm{xc}}(\boldsymbol{r}).
\end{equation}
The first term in the effective potential represents the electron-nuclear potential, the second the Hartree interaction, and the third the exchange-correlation potential.
The end result of the KS approach is that we can solve the ground state density for the interacting system by first solving the Schr\"{o}dinger equation for a single particle in an effective potential, 
using the set of KS equations defined as:
\begin{equation}
\left[\frac{\hat{p}^{2}}{2m} + V_{\mathrm{s}}(\boldsymbol{r})\right]\phi_{i}(\boldsymbol{r}) = \varepsilon_{i}\phi_{i}(\boldsymbol{r}).
\end{equation}
The solutions of these KS equations give a set of KS wavefunctions and energy levels labelled by $i$.   
These states are filled up by the set of available electrons and the density can be constructed from these non-interacting wavefunctions as $n(\boldsymbol{r}) = \sum|\phi_{i}(\boldsymbol{r})|^{2}$, where the sum runs over the occupied states. 

Although we now have a set of single-particle problems to solve, we don't actually know the potential.
Our ignorance has two aspects: the first is the functional form of $V_{\mathrm{xc}}$, the second is 
that the Hartree potential $V_{\mathrm{H}}$ and the exchange correlation potential $V_{\mathrm{xc}}$ both depend on $n(\boldsymbol{r})$,  the solution to the problem, which in turn is constructed from $\phi_{i}(\boldsymbol{r})$. 
The first problem can be addressed by approximating $V_{xc}$, using for example the local density approximation (LDA)\cite{KohnSham1965}
or some more sophisticated approach.\cite{Becke1993,Perdew1996,Burke2012}
We then solve the equations iteratively: (i) an initial guess is first  made of $n(\boldsymbol{r})$ and the potential $V_{\mathrm{s}}$ is computed; (ii) the single-particle wavefunctions are computed; (iii) the set of occupied $\phi_{i}$ is used to compute $n(\boldsymbol{r})$, and then we start again by returning to step (i) and recalculating the potential. This procedure is repeated until some suitable convergence criterion is reached. 

Central to the use of DFT is the Born-Oppenheimer (BO) approximation.
This approximation  makes use of the fact that atomic nuclei are much heavier than electrons, and therefore that the energy scale for the electronic part of the wavefunction is significantly larger than that of the nuclear part of the wavefunction, allowing these to be separated.
The approximation therefore allows us to treat the atomic positions as parameters of an electron-only Hamiltonian that obeys a Schr\"{o}dinger equation.
The Born-Oppenheimer approximation basically clamps the nuclei at given fixed positions  and the appropriate electron wavefunction is the one corresponding to this clamped nuclear configuration.
However, nothing tells us whether  the configuration of nuclei we  specify is an equilibrium one. Since one of our main purposes in using DFT is finding the details of the nuclear structure that accommodates the muon, we must repeat our calculation of the electronic density for different nuclear structures until a global minimum in the ground state energy has been found. 
To do this, we allow the nuclei to move under the effect of forces, until these forces become small enough that we judge the system is at equilibrium. This process is called a {\it geometry optimization} (or a {\it relaxation} of the structure).

The BO approximation will be important in applying DFT techniques to a crystal containing a muon impurity. The muon is treated as a nucleus (a reduced-mass hydrogen) which, like all nuclei within BO, is clamped into position. The fact that the muon is 1/9 the mass of a proton, and therefore somewhere between a nucleus and an electron, ultimately limits the applicability of the technique, as we discuss in Section~\ref{sec:quantumeffects}.

\subsection{Adding the muon}

The DFT+$\mu$ technique is based around using DFT to optimize the geometry of a material along with a muon impurity, using methods developed for first-principle calculations of defect states.\cite{Freysoldt2014} In practice we can start by
randomly assigning an initial muon site
in a unit cell of the target material and relaxing the structure by allowing both the muon and
the atoms in the crystal to adjust their positions via repeated iteration until
the forces on them are reduced below the threshold. We then evaluate the total energy of the final
relaxed configuration. 
For crystalline materials, we generally use DFT codes that assume periodic boundary conditions. Since the muons in a $\mu$SR experiment are implanted in the ultra-dilute limit, to ensure the boundary conditions never result in muon-muon interactions, we usually 
specify the structure via a 
supercell, containing a number of unit cells and a single muon.
The charge state of the muon is determined by the overall charge of the defect [$+1$ for diamagnetic and neutral (zero) for paramagnetic states] which is fixed initially; note that for the diamagnetic case a uniform, smeared-out Hartree-like compensating charge is added to the supercell ensure overall neutrality,\cite{Leslie1985,Makov1995,VandeWalle1994}
and the final charged state of the impurity is to be determined by the final relaxed configuration.

The geometry optimization is repeated for several other randomly-chosen initial
muon sites, with the lowest total energy of the relaxed structure yielding the
most likely muon site, as well as providing an estimate for the structural
distortion the muon introduces. Often this procedure will lead to a range of candidate muon sites being produced. It is then necessary to form these into clusters of roughly-equivalent sites, which is most efficiently done by making use of the underlying symmetry of the unperturbed material's structure, along with some assumptions about the nature of the distortion and the energy ranges of sites that are likely to be realised. 
The method has proved extremely successful,
providing results which are in quantitative agreement for cases in which the
muon site can be independently verified (for example, ionic fluorides
provide a valuable test-bed for evaluating the method because of the
quantum coherent oscillations produced in  F--$\mu$--F states,\cite{Brewer1986} see Section~\ref{sec:fluorides}). 

Since this procedure is rather well defined for most systems, it is possible to create simple software solutions that allow it to be carried out automatically. The first calculations\cite{Moller2013,Bernardini2013} utilised bespoke programs calling the open-source electronic structure code QUANTUM ESPRESSO.\cite{Giannozzi2009} Recently some examples have started to become available for general use. 
 MuFinder\cite{Huddart2022b} is a tool that enables users to carry out muon-site  calculations via a simple graphical user interface (GUI). The procedure for calculating muon sites, by generating initial muon positions, relaxing the structures, and then clustering and analysing the resulting candidate sites, can be done entirely within the GUI. 
 The software was originally configured to make use of the plane wave electronic structure code {\sc Castep}.\cite{clark2005}
 Once candidate muon sites have been determined, the local magnetic field at the muon site can then also be computed within the program, making use  of the {\it  Magnetic structure and mUon Embedding Site Refinement}
 (MuESR)
 software,\cite{bonfa2018}  allowing the connection between the muon sites obtained and experiment to be made. Alternative software tools are also available, including one based on
a combination of ab initio random structure searching (AIRSS) and machine learning. \cite{Liborio2018} Here, AIRSS is a general scheme for   randomly generating possible structures a system can adopt and then introduce
biases based on chemical, experimental and/or symmetry grounds.\cite{Pickard_2011}

The method described so far can be quite computationally costly, owing to the need for many iterations of the optimization procedure to relax the structure. In view of this, an alternative scheme is the unperturbed electrostatic potential (UEP) method, in which  the charge density of the host material obtained from a DFT calculation is used, unperturbed by the presence of the muon. This provides an estimate the Coulomb force acting on each of the $\mu^{+}$ in the initial muon-containing structures. This potential is a direct output of the DFT calculation, being the opposite of the Coulomb potential experienced by electrons, and its minimum represents the candidate site where the muon feels zero force. Although approximate, the UEP method is extremely fast and works reasonably well in materials in which atomic displacements don't play a major role in stabilising a diamagnetic muon site.\cite{Sturniolo2020} 

To further reduce computational cost,  semi-empirical methods can be employed, which rely on simplifying approximations and parameterizations to allow more efficient computations. 
Hartree-Fock (HF) theory \cite{lancaster2014quantum} was much-used before DFT became established. 
For speeding up HF calculations, a useful series of semi-empirical methods were developed in which a set of empirical parameters is derived for each chemical element treated by the method. 
This is done by optimizing against a training set of molecular data, which includes properties such as geometry, ionization energy, and electric dipole moment, and is generally limited to light atoms and molecular materials.\cite{blundell2022}
In view of the success of the  efficient HF semi-empirical approach, attempts were made to apply similar semi-empirical ideas to the DFT calculation framework. 
This led to the method known as density functional tight binding (DFTB), which focuses on parametrizing the interactions between pairs of atoms.\cite{Sturniolo2019}
The DFTB method is relatively fast, since each stage in the geometry relaxation only requires a single diagonalization of the energy matrix, rather than the iterative self-consistent-field loop of a usual DFT calculation. 
Although  the semi-empirical
 methods are useful for rapidly obtaining  geometries for a subset of problems, such as muoniated molecular radicals, they are not able to produce accurate spin structures.
Thus, in order to produce a reliable spin distribution and corresponding hyperfine parameters, it is necessary to follow on with a single-point DFT calculation after the semi-empirical geometry optimization.\cite{blundell2022}

\section{Examples}

The location of the muon site in a solid is chiefly determined by the electrostatic interactions of the positively-charged muon and its surroundings. In metals, we expect to find bare muons (often called diamagnetic muons), uncoupled from unpaired electron density. In insulators, there is also the possibility of forming muonium (Mu): a bound state of an electron and muon with a binding energy of 13.54~eV. However, our interest below is mostly in bare muons. We expect a stopped bare muon to be {\it self trapped} by its own local distortion to the lattice,\cite{blundell2022} such that (classically) it sits at the bottom of an electrostatic potential well.  
The energy scales chiefly at play determining muon sites are therefore  electrostatic ones of order 0.1--10~eV. This large range of  scales, which is often reflected in the energy differences between candidate muon-containing structures found using DFT,  follows from that found in the cohesive energies of different types of solid, from molecular crystals (of order 0.1~eV/molecule), through metals (1~eV/atom), to insulators (10~eV/atom).\cite{Ziman1972} Such scales are well described by electronic-structure methods such as DFT. Its worth noting that magnetic interactions, whose energy scales are typically determined by interactions of order tens of millivolts and below, are rather smaller still. So while the electronic state of a material, and the consequent distribution of electron density, will likely affect the muon-stopping state, the magnetic state of a material (and the consequent distribution of electron spin density) is  unlikely to determine the muon site.

\subsection{Fluorides}\label{sec:fluorides}
Ionic fluorides were one of the first systems\cite{Moller2013,Bernardini2013} to be studied using DFT+$\mu$.  This is because the muon site in fluoride systems can be identified using the characteristic experimental signature of an F--$\mu$--F state.\cite{Brewer1986}  Even though there is no electronic
magnetism in many of these compounds, a clear and rather complex precession
signal was observed. has some rather special properties:
(i) fluorine is extremely electronegative and thus attractive
to the positively-charged muon, (ii) F$^-$ has a very small ionic radius (so that the muon sits
very close), (iii) the fluorine nucleus
has spin one-half (so that any relaxation retains
a simple, yet characteristic, time dependence) and (iv) the fluorine
nucleus
has a
large nuclear moment, associated with an isotope which occurs with
100\% abundance (so that all states behave the same).  It was
recognised that the effect could be explained if a muon stops between
two fluorine ions to form what is known as an F--$\mu$--F state.
This state can be described by the Hamiltonian ${\cal H}$
given by
\begin{equation}
{\cal H}= {\cal H}_{\mu {\rm F}} + {\cal H}_{{\rm FF}},
\label{eq:hamiltonianFmuF}
\end{equation}
the sum of two terms
expressing the dipolar interaction between the muon and the fluorine
nuclei
\begin{equation}
{\cal H}_{\mu{\rm F}}=  {\mu_0 \over 4\pi} 
\sum_{i=1}^2 
{\hbar^2 \gamma_\mu \gamma_{\rm F}  \over r_i^3}
\left[
  \boldsymbol{S}_\mu\cdot\boldsymbol{S}_{{\rm F}i} - { 3(\boldsymbol{S}_\mu\cdot\boldsymbol{r}_i) 
                         (\boldsymbol{S}_{{\rm F}i}\cdot\boldsymbol{r}_i) \over r_i^2}
\right],
\label{eq:hamiltonianFmuF1}
\end{equation}
and the dipolar interaction between the fluorine nuclei
\begin{equation}
{\cal H}_{{\rm FF}}=  {\mu_0 \over 4\pi} 
{\hbar^2 \gamma_{\rm F}^2 \over r_{\rm FF}^3}
\left[
  \boldsymbol{S}_{{\rm F}1}\cdot\boldsymbol{S}_{{\rm F}2} - { 3(\boldsymbol{S}_{{\rm F}1}\cdot\boldsymbol{r}_{\rm FF}) 
                         (\boldsymbol{S}_{{\rm F}2}\cdot\boldsymbol{r}_{\rm FF}) \over r_{\rm
      FF}^2}
\right].
\label{eq:hamiltonianFmuF2}
\end{equation} 
If we ignore the interaction between the two
fluorine moments, then eigenvalues are $-1$,
$\frac{1-\sqrt{3}}{2}$, $0$, and $\frac{1+\sqrt{3}}{2}$, all in
units of $\hbar\omega_{\rm d}=  \mu_0 \hbar \gamma_\mu \gamma_{\rm F} /( 4\pi r^3) $ and
twice-repeated, so the energy levels are four doublets.
The polycrystalline average polarization is then given by an analytical form    
\begin{equation}
\langle  P_z(t) \rangle =  \frac{1}{2}
+  \frac{1}{6} \sum_{i=1}^3 a_i \cos(\alpha_i\omega_{\rm d} t),
\label{eq:fmufgeneral}
\end{equation}
    with 
    $a_1=1$, 
    $a_2=1-\frac{1}{\sqrt{3}}$,
    $a_3=1+\frac{1}{\sqrt{3}}$,
    $\alpha_1=1$,
    $\alpha_2=\frac{3-\sqrt{3}}{2}$, and
    $\alpha_3=\frac{3+\sqrt{3}}{2}$. 
If the interaction between the two fluorine moments is also
included then the factors in this expression change a little
  bit (and the expression doesn't look as pretty).
Nevertheless, the form of the oscillatory signal provides information concerning the nature of the interaction (whether with one fluorine, or two, and whether the F--$\mu$--F bond is straight or bent) and the timescale of the oscillatory signal is set by $\omega_{\rm d}$, and hence by $r$, so that the muon-fluorine bond length can be accurately measured. This provides crucial quantitative information on the muon site and its local distortion, allowing a rigorous test of the accuracy of the DFT$+\mu$ calculations.

The oscillations in the muon polarization described in eqn~\ref{eq:fmufgeneral} have been observed experimentally in many inorganic fluorides,\cite{Brewer1986,Noakes1993} fluoropolymers,\cite{Pratt2003,Nishiyama2003,Lancaster2009} and fluorine-containing molecular magnets \cite{Lancaster2007}. 
DFT$+\mu$ calculations\cite{Moller2013} in a variety of inorganic fluorides (LiF, NaF, CaF$_2$, BaF$_2$, and CoF$_2$) reveal a linear F--$\mu$--F state with a F--F internuclear distance between 2.31 and 2.36\,\AA, whose range includes 2.36\,\AA, the measured distance of the (FHF)$^-$ molecular bifluoride ion in vacuum.\cite{Kawaguchi1987}
The bifluoride ion has various vibrational modes (e.g.\ a bending mode of frequency around 1290\,cm$^{-1}$), but the calculated modes for F--$\mu$--F are around a factor of three larger\cite{Moller2013} (due to the muon mass being approximately one ninth of the proton mass).  This results in the muon in a F--$\mu$--F state having a zero-point energy of $\approx 0.8$~eV, considerably larger than the value of $\approx 0.3$~eV for the proton in
the bifluoride ion.\cite{Moller2013}. We will return to this point in Section~\ref{sec:quantumeffects}.

\begin{figure}
\includegraphics[width=\columnwidth]{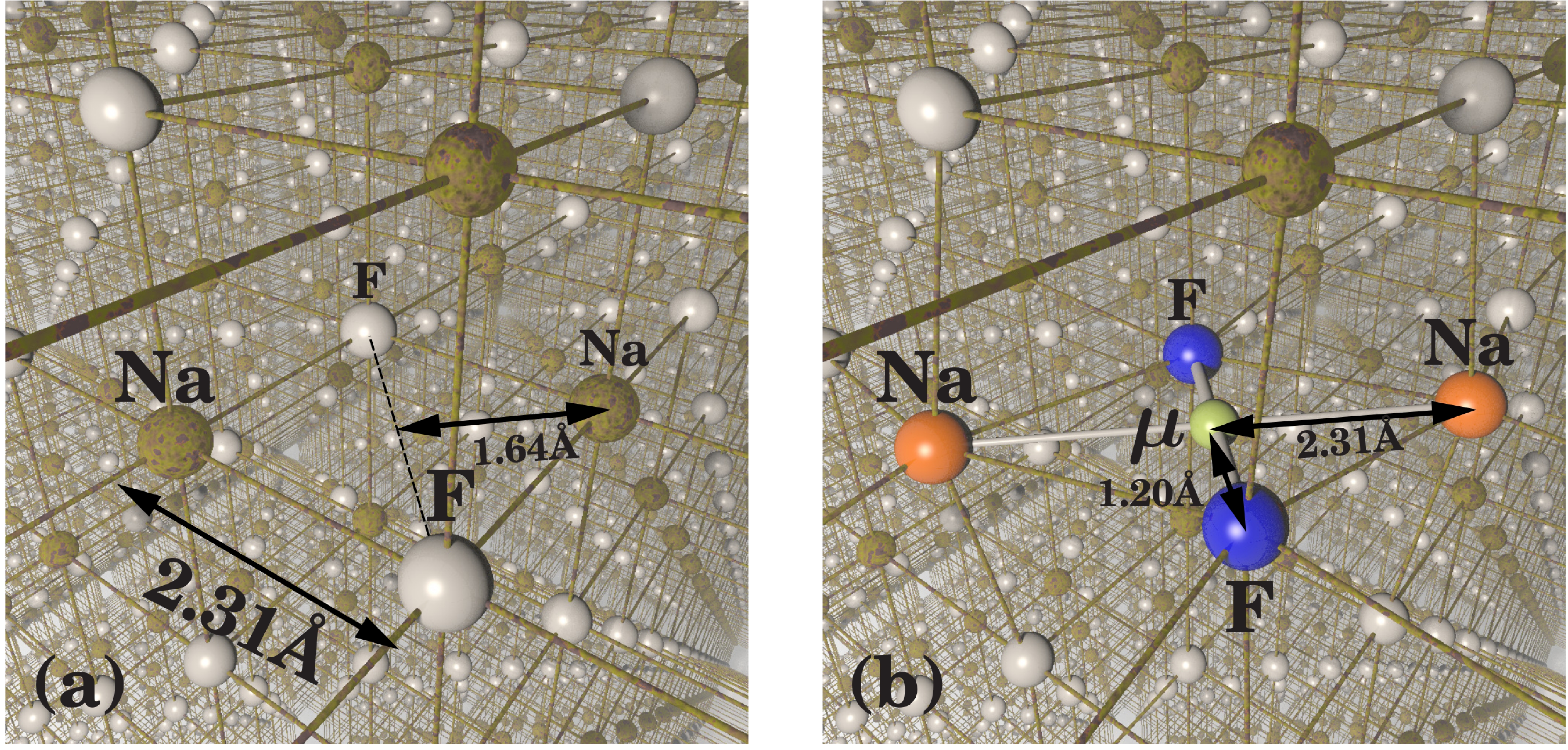}
\caption{\label{fig:naf} 
(a) The crystal structure in NaF. (b) After implantation of the muon, two nearby fluorine ions are pulled in and two nearby sodium ions are pushed out. [Adapted from J. Wilkinson and S. J. Blundell, Phys.~Rev.~Lett. {\bf 125}, 087201 (2020).\cite{Wilkinson2020}]
Copyright 2020 American Physical Society.}
\end{figure}

The rigidity of the F--$\mu$--F bond will introduce a distortion into the crystal. For example, in CaF$_2$, the F--$\mu$ distance is found by experiment to be 1.172\,\AA\ (with the DFT$+\mu$ calculation giving 1.134\,\AA), but this value is much lower than $a/4=1.362$\,\AA, the distance between the muon site and the fluorine ion {\it in its undistorted position}, demonstrating a significant distortion.\cite{Wilkinson2020}  The same effect in NaF is illustrated in Fig.~\ref{fig:naf}.  In the undistorted structure, each ion sits 1.64\,\AA\ from the site of the muon; following implantation, the distance from the muon to the nearest-neighbour fluorine (sodiuum) ion becomes 1.20\,\AA\ (2.31\,\AA).  The results of these DFT$+\mu$ calculations are in excellent agreement with experimental results for which the complex precession signals due to the dipolar couplings with all nearby nuclei can now be calculated and high-statistics measurements have now been performed.\cite{Wilkinson2020}

\begin{figure}
    \centering
\includegraphics[width=\columnwidth]{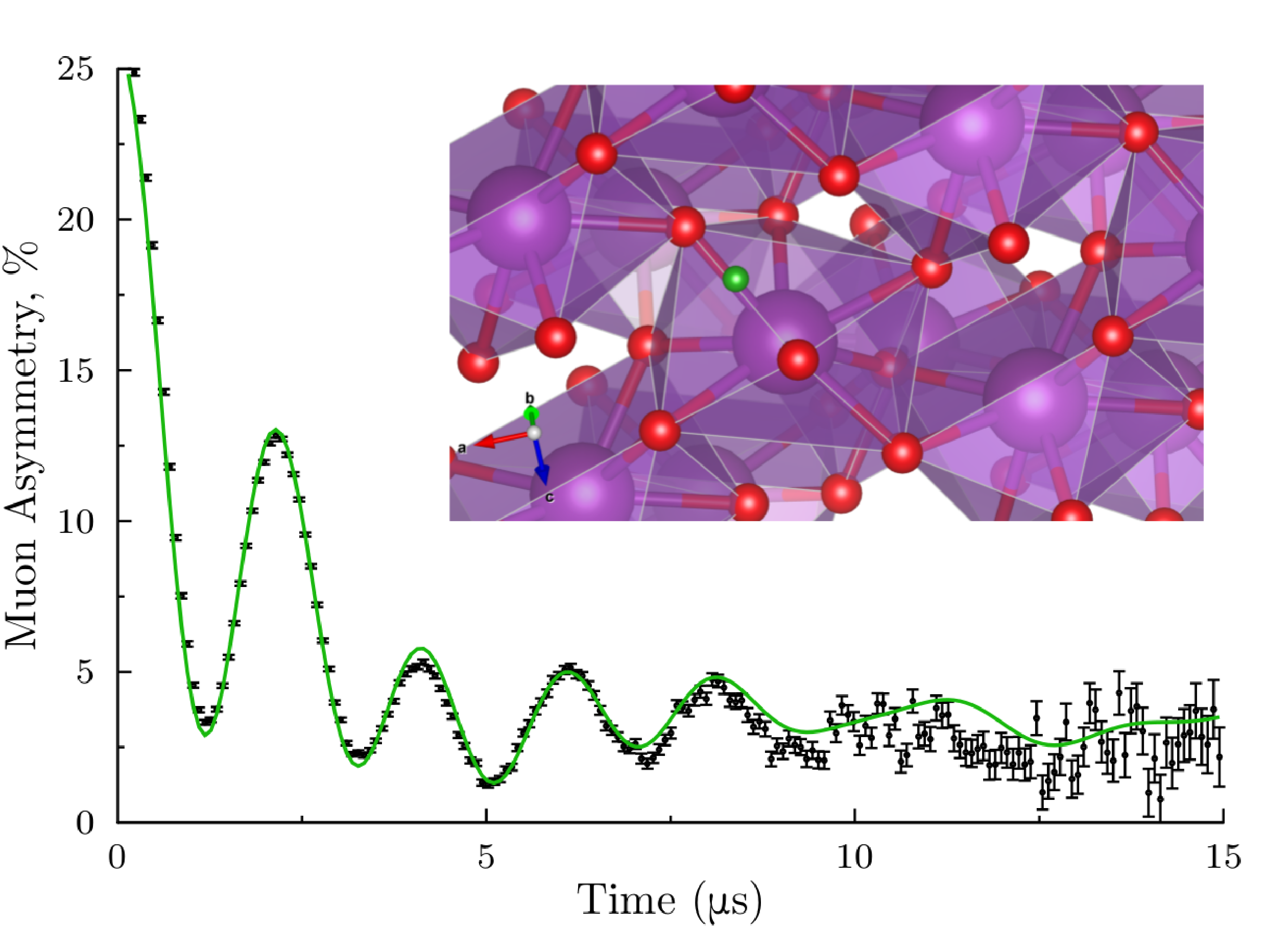}
    \caption{$\mu$SR experimental data on YF$_3$ with the fitted muon polarization using the muon site obtained by DFT$+\mu$. This site is displayed on the YF$_3$ crystal structure in the inset.
    [Adapted from J. Wilkinson {\sl et al.} Phys. Rev. B {\bf 104}, L220409 (2021).\cite{Wilkinson2021}]
    Copyright 2021 American Physical Society.}
    \label{fig:yf3}
\end{figure}

This approach has now been extended to other fluorides, including YF$_3$ (see Fig.~\ref{fig:yf3}) and $\alpha$-PbF$_2$; in the latter compound the $\mu$SR data and analysis provide evidence for a Mu$^-=\mu^+\mbox{e}^-\mbox{e}^-$ stopping site in an anion Frenkel defect.\cite{Wilkinson2021}  Although fluorine is the most ideal ion for exploring these effects, it is not the only one.  $\mu$SR data obtained on V$_3$Si shows evidence for quantum coherent oscillations which can be related to a V--$\mu$--V state.\cite{Bonfa2022}  These experiments highlight
 the extreme sensitivity of the entangled states to the local structural and electronic environments
 which, in the case of these A15 compounds containing V (and also Nb), emerges through the quadrupolar interaction with the electric field gradient.  This demonstrates that positive muons, usually thought of as a purely magnetic probe, can also be deployed as quantum sensors to measure structural and charge-related phenomena.\cite{Bonfa2022}

\subsection{Magnetism}
Magnetism remains a key research area for $\mu$SR and many applications of the DFT+$\mu$ techniques have concentrated on systems that are magnetically ordered. Here a knowledge of the muon site tells us which magnetic fields are causing the time-evolution of the muon's spin, potentially giving us access to more information on the local magnetic field distribution and its dynamics. 

A simple, topical example is found in materials adopting the double-perovskite structure  exemplified by antiferromagnetic Sr$_{2}$FeOsO$_{6}$.\cite{Williams_2016} In this system the muon site is found from DFT to be approximately 1\,\AA\ from an oxygen atom in the basal plan of the oxygen octahedra surrounding the Os ions, validating the rule of thumb that such sites close to O$^{2-}$ ions are most likely in oxides. In this case the site is consistent with the dipole-field map computed for the known magnetic structure, giving confidence that the DFT-derived site is likely the one realised.   
Other ionic systems in which DFT$+\mu$ calculations show that the muon occupies a site close to the anion include the honeycomb system\cite{Lang2016} $\alpha$-RuCl$_3$, and the spin-Jahn-Teller antiferromagnets\cite{Kirschner2019,Lang2019} CoTi$_2$O$_5$ and FeTi$_2$O$_5$.   
A similar approach applies in the case of the ferromagnet Nd$_2$Fe$_{14}$B, where the muon site is identified [the 8i site (0.6745,0.8838,0)] near the square base of a NdFe$_3$B pyramid, leading to a quantitative measurement (using the observed $\mu$SR precession frequencies) of the moment on the Nd and Fe atoms.\cite{Sugiyama2019}

\begin{figure}
\includegraphics[width=0.7\columnwidth]{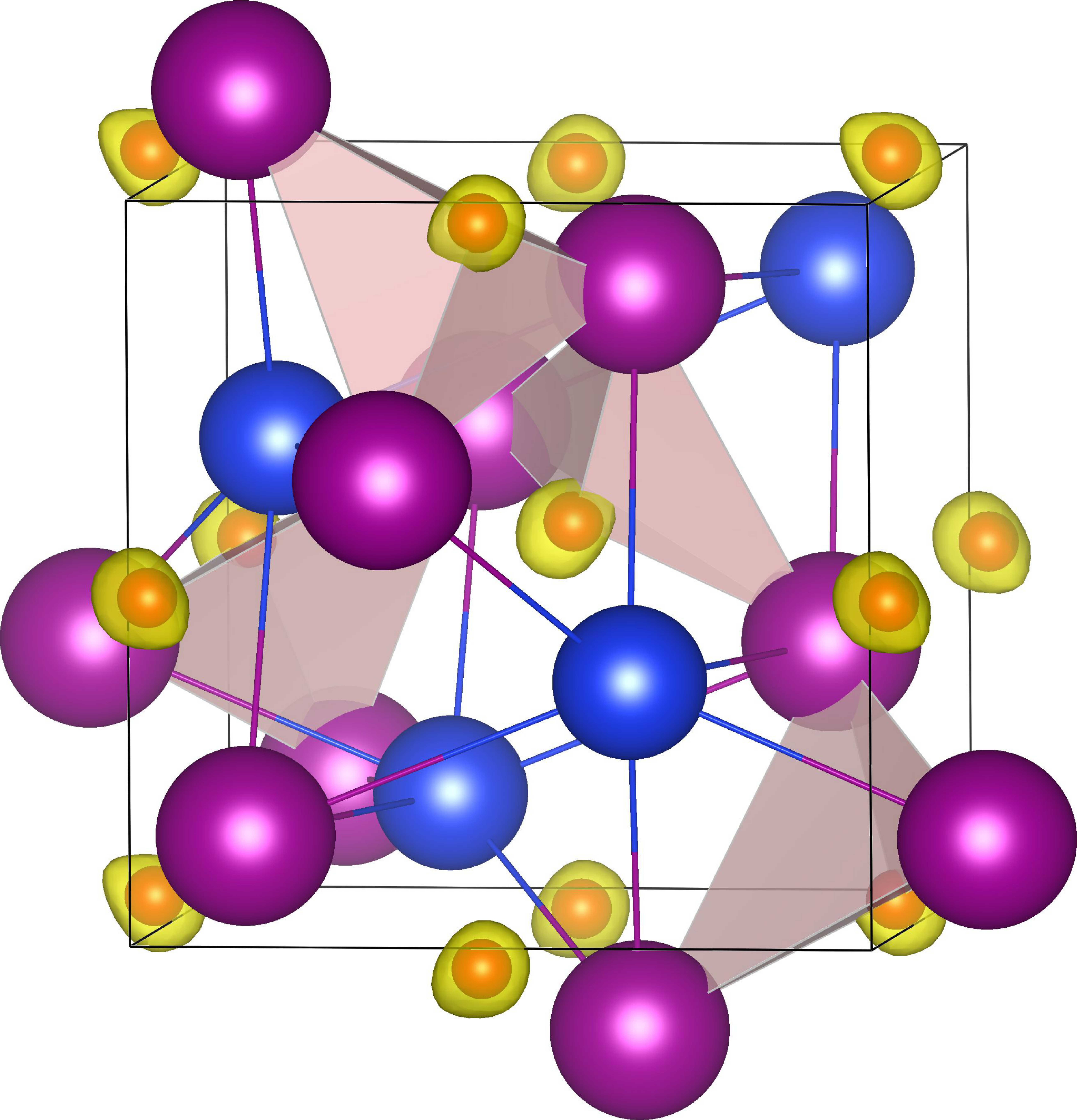}
\caption{\label{fig:mnsi} 
The muon site in MnSi is found experimentally to be at  $(0.532,0.532,0.532)$ (indicated by small circles at crystallographically equivalent sites). These positions are enclosed by the regions where the unperturbed electrostatic potential computed by DFT takes a minimum. [Adapted from A.\ Amato {\it et al.}, Phys.~Rev.~B {\bf 89}, 184425 (2014).\cite{Amato2014}] Copyright 2014 American Physical Society.}
\end{figure}

More stringent tests of DFT-computed sites are found in materials with more complicated, often noncollinear, magnetic structures and/or hyperfine contributions. 
The helimagnet MnSi has been the subject of several illuminating investigations of its properties using muons \cite{Amato2014,Bonfa2015} and provides a good example of site-determination in a more complicated magnetic system, 
 In this material
  the muon site computed using DFT methods (shown in Fig.~\ref{fig:mnsi})
is consistent with the symmetry properties derived from Knight-shift measurements, and also with the very restrictive details of the complicated magnetic structure of the material.\cite{Amato2014} This information proved very useful in providing a complete analysis of the measured $\mu$SR spectra in terms of a magnetic structure and hyperfine interaction,\cite{Amato2014} that did not require the invocation of a more exotic muon stopping state that had been previously suggested.\cite{Storchak2011}

\begin{figure}
\includegraphics[width=0.48\columnwidth]{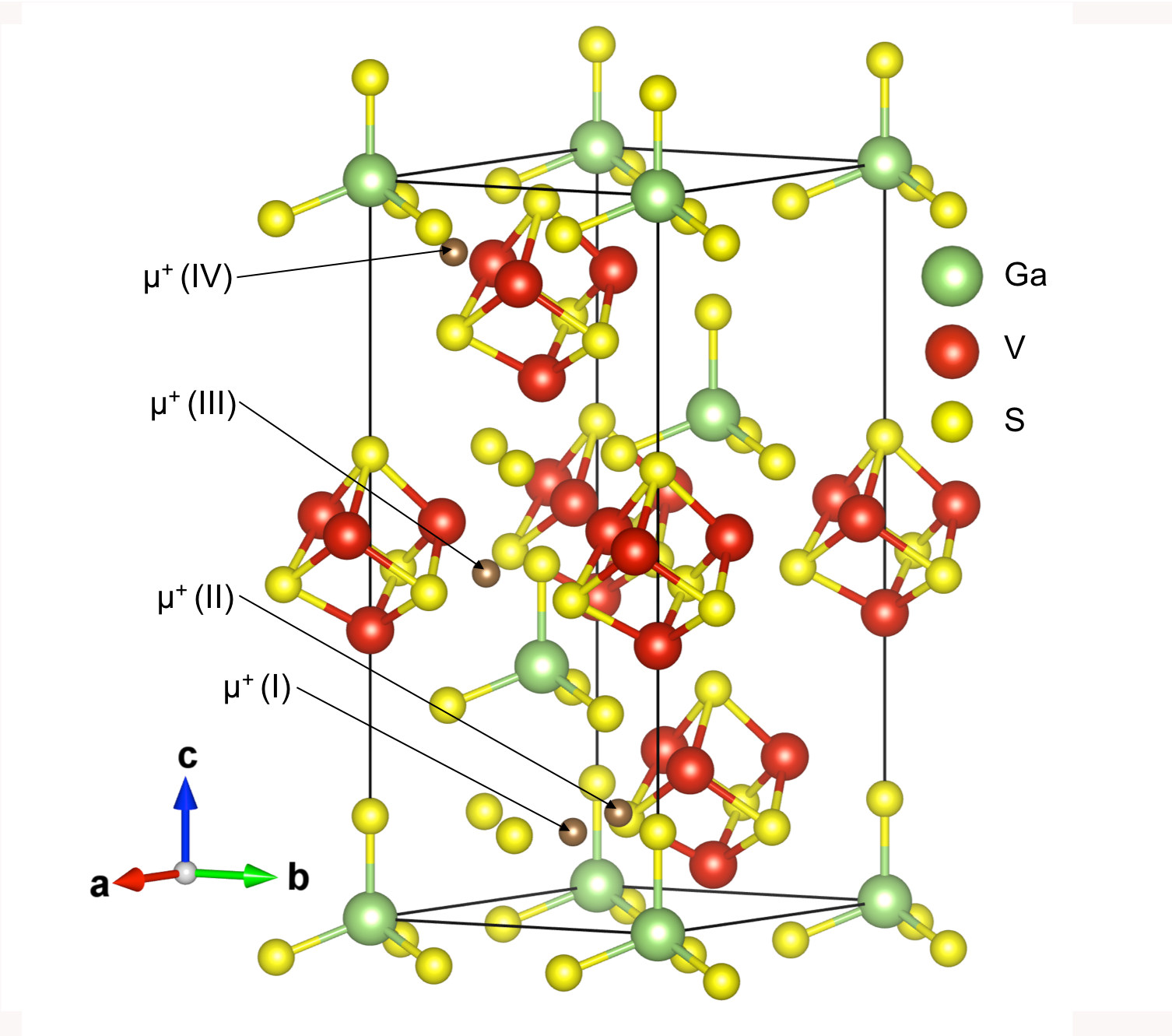}
\includegraphics[width=0.48\columnwidth]{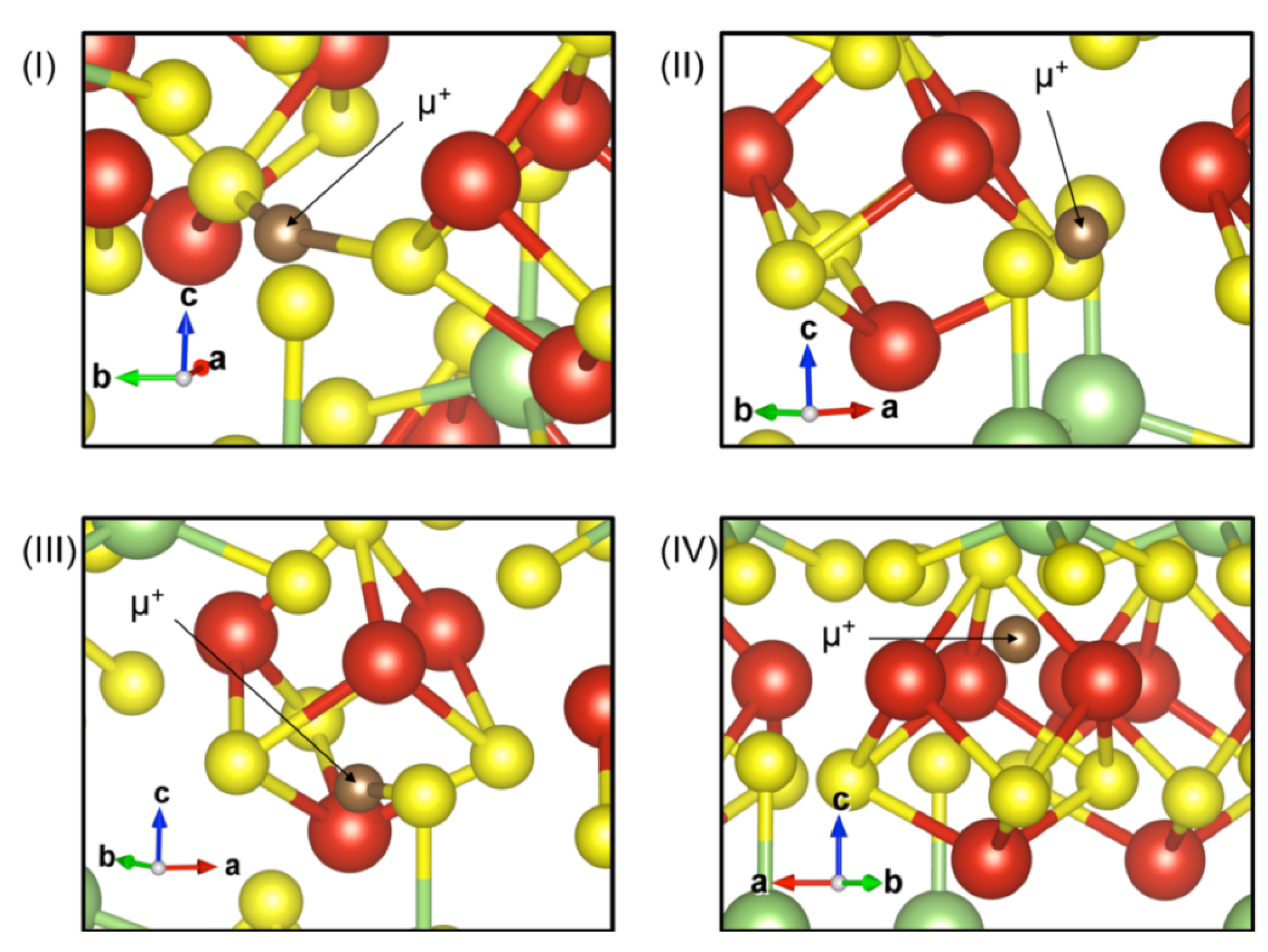}
\caption{\label{fig:gavs1} {\it Left:} the four classes of muon stopping site determined for
GaV$_{4}$S$_{8}$. {\it Right:} local geometry around the muon for each of the four
classes of muon stopping site in GaV$_{4}$S$_{8}$.
The sites are numbered in order of increasing energy.
 [Adapted from K.\ J.\ A.\ Franke {\it et al.}, Phys.~Rev.~B {\bf 98}, 054428 (2018).\cite{Franke2018}] Copyright 2018 American Physical Society.}
\end{figure}

Noncollinear magnetism magnetic structures are especially relevant to the study of skyrmion systems. Skyrmions \cite{Lancaster2019} are vortex-like magnetic excitations found in a growing number of magnetic materials. These include MnSi, which hosts Bloch skyrmions. Another example skyrmion compound, this one hosting N\'eel skyrmions, is
GaV$_{4}X_{8}$ ($X$=S, Se).
Structural relaxations of a supercell of GaV$_{4}$S$_{8}$\cite{Franke2018}  reveal four distinct muon sites [Fig.~\ref{fig:gavs1}].  Three of these (labeled I-III) involve the muon sitting close to a single S atom. A fourth site (site IV) has the muon closer to V atoms and is the highest energy site. In the lowest energy site (site I) the muon sits between two S atoms, in the plane defined by three S atoms within V$_4$S$_4$ units.  The two $\mu^+$--S distances are unequal (1.4\,\AA~ and 2.0\,\AA) with greater electron density found between the muon and the nearest S atom. This site is therefore best described in terms of the muon forming a $\mu^+$--S bond (rather than an S--$\mu^+$--S state), though the presence of a second nearby S atom does seem to stabilize this geometry.
Two further sites involve the muon sitting close to a single S atom. 

Perhaps the most pressing problem with the DFT+$\mu$ method is determining which sites are actually realized given an energy-ordered list of candidate states. This problem is particularly relevant for 
the results of analogous calculations for GaV$_4$Se$_8$. These give stopping sites (labeled $1$ - $4$ in order of ascending energy) that are similar to those calculated for GaV$_4$S$_8$, with three of the four sites involving the muon sitting close to a Se atom (sites $2$ to $4$) and a site in which the muon sits above a face of a V$_4$Se$_4$ unit (site~1). However, the ordering of sites is inverted in the Se case, compared to the S-containing series. In particular, the cube face site (site~1), which corresponds to the highest energy stopping site for GaV$_4$S$_8$, is the lowest energy site for GaV$_4$Se$_8$.  Although the difference in energy ordering is interesting, without access to a method to compute the capture cross section for each site, the interpretation of this information remains an open question.

So far we have seen examples of the commonly  encountered cases of sites near fluorine and near oxygen. 
An interesting case of a system where both sites near oxygens {\it and} fluorines are predicted is barlowite, a
frustrated magnet kagome antiferromagnet with formula [Cu$_{4}$(OH)$_{6}$FBr].\cite{Tustain2022} In this system, DFT suggests 
two distinct classes of muon stopping site. The first 
 localises $\approx 1.0$\,\AA\ away from the oxygen
atoms in the hydroxide groups that connect the Cu$^{2+}$ ions within
the kagome layers. This forms a triangular $\mu$--OH
complex, with the muon-proton distance of 1.54\,\AA\ in the
lowest-energy sites. In the second class, muons 
localise near the fluoride
anions in between the kagome layers, with a $\mu$--F
separation of 1.1\,\AA. These latter sites lie at
substantially higher energies above the lowest energy  sites
in the calculations ($\approx 1$~eV). Although this suggests, on purely
energetic grounds, that the formation of $\mu$--F complexes in
barlowite is unlikely compared to $\mu$--OH. However, the measured spectra in the paramagnetic regime show clear dipole-dipole oscillations consistent with a $\mu$--F complex, implying  that during
the stopping process muons are indeed captured in these latter
potential minima.

The study of H-like defects in semiconductors and insulators using first-principles methods is a mature field,\cite{VandeWalle1994,VandeWalle2000,Freysoldt2014} and understanding the analogous muon-like states is clearly related.\cite{Cox2009,Hirashi2022}
A recent development has come from the suggestion that in certain oxides, such as Cr$_2$O$_3$, the muon can form a charge-neutral complex composed of a muon and an electron polaron.\cite{Dehn2020}  Cr$_2$O$_3$ contains Cr$^{3+}$ (3d$^3$) ions, but the idea is that the $\mu^+$ binds to an oxygen anion but an excess electron localizes on an adjacent Cr ion, changing its charge state to Cr$^{2+}$ (3d$^4$), which is Jahn-Teller active, resulting in a Jahn-Teller polaron.  The existence of this state is supported by first-principles calculations, and an analogous effect has been proposed\cite{Dehn2021} for Fe$_2$O$_3$.  

\subsection{Superconductivity}
An important use of muon in superconducting materials is  the investigation of the vortex lattice in type-II superconductors, where the muon can be used to measure the penetration depth. In such measurements, the muon site is often of secondary importance since the length scale of the vortex state is large on the scale of the underlying unit cell, so the muons effectively probe the whole of the flux lattice, whatever their stopping site in the crystal. However, the physics of unconventional superconductors is inextricably bound up with magnetic phenomena (and sometimes questions of phase separation and coexistence), and so in $\mu^{+}$SR studies of materials showing unconventional superconductivity, it is usually absolutely necessary to have knowledge of the muon stopping state. 

\begin{figure}
\includegraphics[width=0.78\columnwidth]{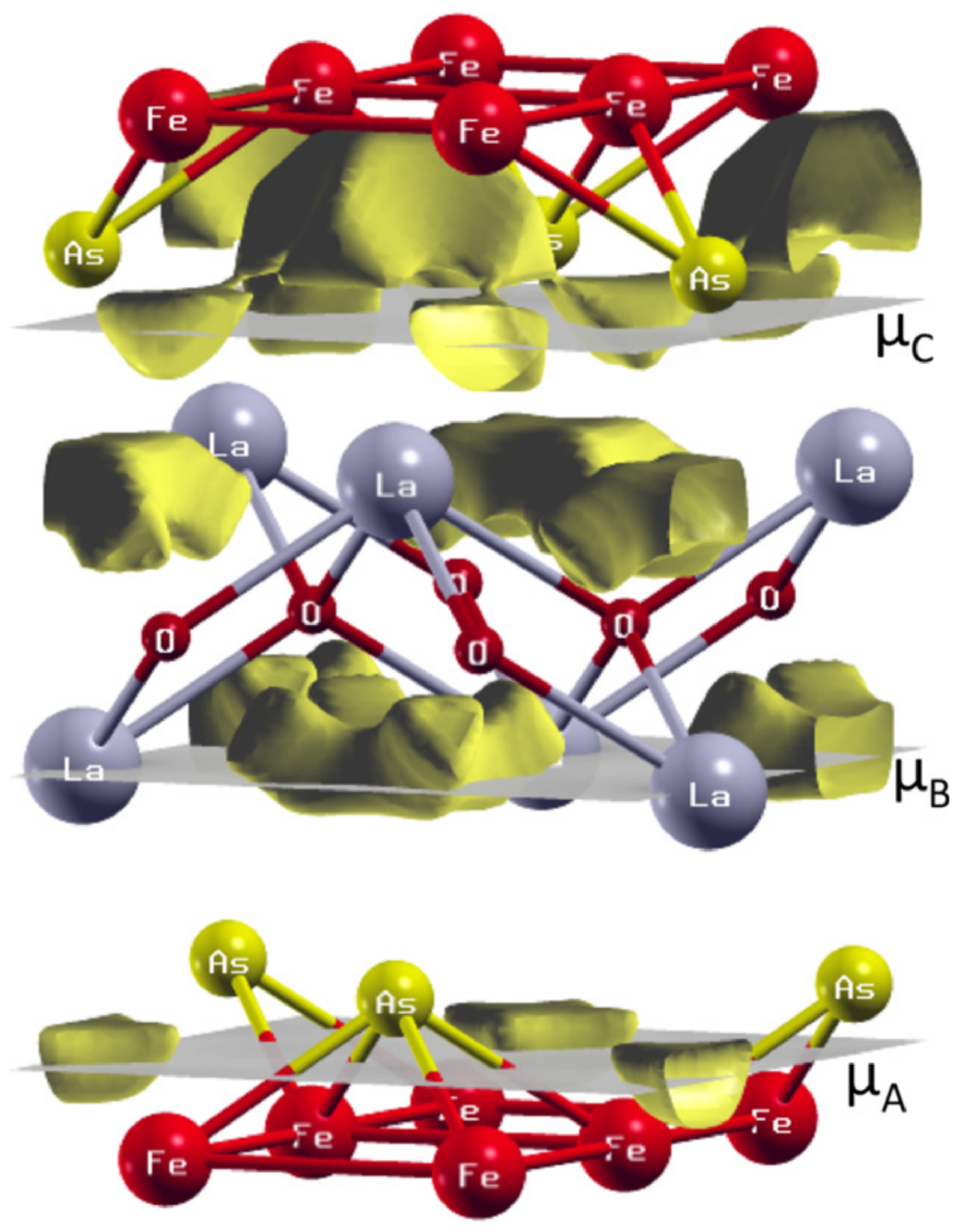}
\caption{\label{fig:pnictides}
Muon sites identified in the LaFeAsO unit cell. Gold-shaded areas represent the volume of zero point displacement for the muon.
[Taken from R. De Renzi {\it et al.}, Supercond.\ Sci.\ Technol.\ {\bf 25}, 084009 (2012).\cite{De_Renzi_2012a}]
Copyright 2012 Institute of Physics Publishing.}
\end{figure}

In the cuprate superconductors, it has been thought for a long time that the muon site will be about 1\,\AA\ away from an oxygen anion.\cite{Brewer1991}  A recent study of muon sites in La$_2$CuO$_4$ has concluded that most muons stop near the apical oxygen, but there are two additional sites that also receive a tiny fraction of muons.\cite{Ramadhan2022}
Iron-based superconductors provide another informative example of an unconventional superconducting series of materials.\cite{De_Renzi_2012a,Cheung2018} 
 The DFT calculation allows a straightforward identification
of the muon site using the UEP approach. In the case of
the 1111 structure this allows the  identification of three sets of potential minima 
 shown in Fig.~\ref{fig:pnictides}.  Similar results have been obtained for\cite{Sundar2023} UTe$_2$.

One area where $\mu$SR has proven important is in attempts to determine instances of time-reversal symmetry breaking (TRSB) in superconducting systems, since this
 can provide a tight constraint on the symmetry of
the superconducting gap.\cite{Huddart2022}
The assignment of TRSB is made through the appearance of spontaneous
magnetic fields found 
using $\mu$SR measurements. (Importantly,
this effect  is absent for most superconductors.) The effects observed are frequently small, and sometimes only seen in muon measurements and so it is natural to ask whether a muon-induced perturbation is a contributing factor to the observations. 
A recent DFT+$\mu$ study of these materials\cite{Huddart2022} suggest that the muon is an innocent probe of these materials. We shall concentrate on 
one candidate TRSB material: the layered perovskite 
superconductor Sr$_{2}$RuO$_{4}$.
\begin{figure}
\includegraphics[width=\columnwidth]{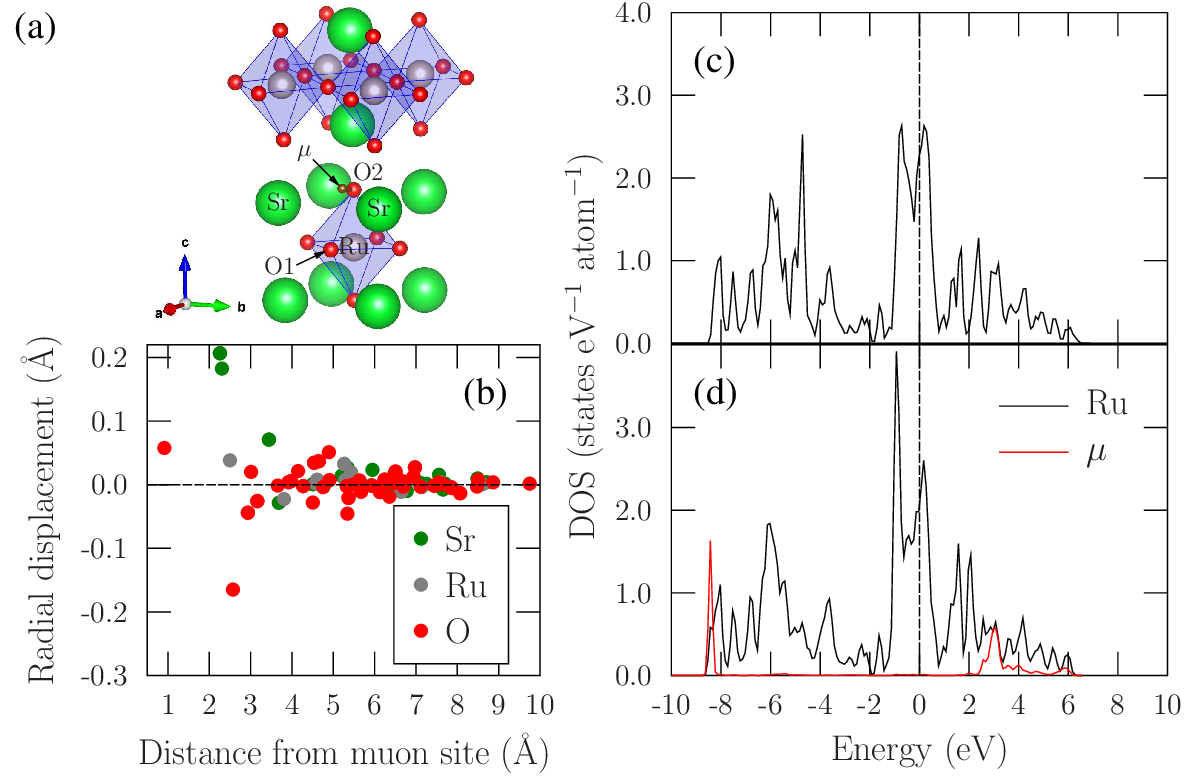}
\caption{\label{fig:SRO} The lowest-energy muon site in Sr$_{2}$RuO$_{4}$. (a) Local
geometry of the muon site. (b) Radial displacements of atoms as a
function of their distances from the muon site. PDOS for the
Ru atom closest to the muon site for the structures (c) without a
muon and (d) with a muon. Energies are given with respect to the
Fermi energy. [Adapted from B.~M.\ Huddart {\it et al.}, Phys.\ Rev.\ Lett.\ {\bf 127}, 237002 (2021).\cite{Huddart2022}] Copyright 2021 American Institute of Physics.}
\end{figure}

In the lowest-energy muon site 
found for Sr$_{2}$RuO$_{4}$, the muon is
bonded to an oxygen (O2) with bond distance 0.973\,\AA, consistent with muon sites in
other oxides including high-temperature superconducting
cuprates  and pyrochlores.  The small muon-induced
displacement vanishes rapidly as a function of distance
from the muon site, such that significant distortions are
observed only for atoms within 6\,\AA\ of the muon site.
We can therefore conclude that there is little structural distortion, 
but what about a distortion to the physics at the next largest energy scale: the electronic structure?
It is found that the dominant contribution to the electronic density of states (DOS) close to the Fermi
energy is that from the Ru atoms. The effect of muon
implantation on the projected DOS of the Ru atom closest to the
muon site is shown in Figs.~\ref{fig:SRO}. There is an 
increase in the DOS at around 1~eV below the
Fermi energy caused by small changes in the splitting of
the Ru 4d$_{zy}$ and 4d$_{zx}$ states at the Fermi level, which are not
observed for Ru atoms further away from muon. However, after
summing the d-state contributions from all of the Ru ions in
the supercell, the small state splitting is no longer resolvable.
The partial density of states corresponding to the muon itself
lies around 8~eV below the Fermi energy. Since this is very large interval in energy, it is highly unlikely the muon has any effect on the electrons near the Fermi energy and so
these results suggest that the implanted muon does
not have a significant effect on  Sr$_{2}$RuO$_{4}$. A similar conclusion was reached for other TRSB systems investigated in this study.\cite{Huddart2022}

\subsection{Molecular systems}

\begin{figure*}
\includegraphics[width=1.9\columnwidth]{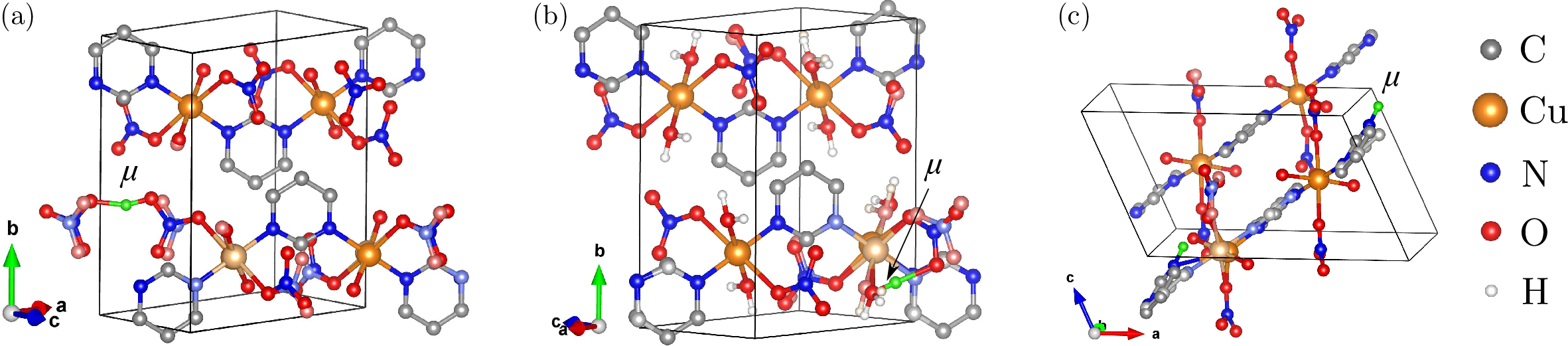}
\caption{\label{fig:staggered} Low-energy muon sites in Cu-PM. Lighter spheres represent the ionic positions in the unit cell without the muon. H atoms have been omitted for clarity where appropriate. (a) The nitrate site. (b) The H$_{2}$O site. (c) The N(pym) site.
[Taken from B.~M.\ Huddart {\it et al.}, Phys.\ Rev.\ B {\bf 103}, L060405 (2021).\cite{Huddart2021}] Copyright 2021 American Institute of Physics.}
\end{figure*}

Muons are routinely used to elucidate the properties of molecular materials. These systems are attractive owing to their tunability, but their chemical complexity regularly leads to questions about the nature of the muon stopping state. These materials are rather soft in comparison to their inorganic counterparts, and so we might expect that the muon-induced distortion could present a relatively large perturbation to the electronic or magnetic properties of the system. 

One class of these materials are coordination polymer magnets, where a magnetic ion is linked via molecular ligands to form a low-dimensional magnet. These are of interest since the low-dimensional character often suppresses long-range magnetic order, reducing the size of the response of conventional measurement techniques owing to the small moment size and small change in entropy upon ordering. In these systems muons are often able to detect transitions that prove invisible to magnetic susceptibility or specific heat. 

The staggered molecular spin
chain [pym-Cu(NO$_{3}$)$_{2}$(H$_{2}$O)$_{2}$] (pym = pyrimidine), known as Cu-PM
is a good  example.\cite{Huddart2021} Here magnetic order was detected using muons at $T=0.23$~K. 
Three distinct classes
of muon stopping site were determined [Fig.~\ref{fig:staggered}]. Sites where
the muon sits around 1\,\AA\ from an O atom in a nitrate group
[Fig. \ref{fig:staggered}(a)] or H$_{2}$O ligand [Fig.~\ref{fig:staggered}(b)] are the lowest and second
lowest energy classes of sites, respectively. We also find candidate
sites where the muon sits 1.0~Å from an N atom in a pym
ligand [Fig.~\ref{fig:staggered}(c)], which are substantially higher in energy and
result in larger local distortions to the crystal structure. These
sites can be mapped to features in the ZF spectra by considering
the dipolar fields resulting from candidate antiferromagnetic
structures. From dipolar field calculations one obtains
fields of 9–40 mT/$\mu_{\mathrm{Cu}}$ for the nitrate site, 57–63 mT/$\mu_{\mathrm{Cu}}$
for the H$_{2}$O site, and 93–99~mT/$\mu_{\mathrm{Cu}}$ for the N(pym) site,
where $\mu_{\mathrm{Cu}}$ is the ordered moment of the Cu$^{2+}$ ions in Bohr magnetons. The relative size of the calculated
fields for the H$_{2}$O and nitrate sites is consistent with the ratio
between observed frequencies measured in the ordered regime. This assignment
gives an estimate $\mu_{\mathrm{Cu}}$ $\approx$ 0.38$\mu_{\mathrm{B}}$ for the ordered moment.\cite{Huddart2021} 

\begin{figure}
\includegraphics[width=0.9\columnwidth]{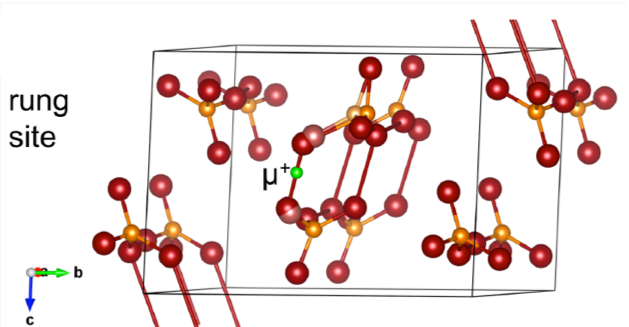}
\caption{\label{fig:ladder}  Example muon site in the molecular spin ladder (Hpip)$_{2}$CuBr$_{4}$. [Taken from T.~Lancaster {\it et al.}, New.~J.~Phys.~{\bf 20}, 103002 (2018).\cite{Lancaster_2018}] Copyright 2018 Institute of Physics Publishing.}
\end{figure}

There have been claims,\cite{eggert1992} supported by some experimental evidence,\cite{chakhalian2003} that a muon implanted in some spin chains could form an unusual spin-singlet state, not unlike that found in the Kondo effect.
However, results on materials such as Cu-PM and the linear chain Cu(pyz)(NO$_{3}$)$_{2}$, \cite{Xiao2015} do not provide evidence for the realization of these states. 
An  example where a substantial muon-induced perturbation seems to be present is the molecular spin-ladder material (Hpip)$_{2}$CuBr$_{4}$,\cite{Lancaster_2018} where the muon forms states based on Br-- $\mu$--Br bonds. This state causes a sizeable local distortion to the atoms neighbouring the muon and consequently to the electronic structure (Fig.~\ref{fig:ladder}). Although it might be feared that this would prevent the muon from faithfully measuring the properties of the material, this seems not to be the case as the magnetic phase diagram is determined by physics on a length scale that is very long compared to the muon and its distortion. As a result the transitions in the material are observed  with $\mu$SR at the fields and temperatures found using other techniques.

An example of a rather different structure is the organic radical 2-(4,5,6,7-tetrafluorobenzimidazol-2-yl)-4,4,5,5-tetramethyl-4,5-dihydro-1$H$-imidazole-3-oxide-1-oxyl (F4BImNN)\cite{blundell2013} which forms hydrogen-bonded chains  and exhibits one-dimensional
ferromagnetic exchange. Here three energetically similar sites are predicted
(i)  where the muon forms a covalent bond with the oxygen in the
nitronyl nitroxide group (see Fig. 4), (ii) 
where the muon forms a covalent bond with the vacant nitrogen
and hydrogen-bonds to one oxygen in the nitronyl nitroxide group (this is the lowest energy site),
and (iii) where the muon bonds to the other
oxygen in the nitronyl nitroxide group. Interestingly, despite the presence of fluorine in this sytem, an F--$\mu$--F state is not predicted.

\subsection{Muon-induced distortions}
This example of a spin-ladder material from the previous section is a case where the muon-induced distortion likely leads to a probe state comprising both the muon and its surrounding perturbation that is nevertheless sensitive to the intrinsic properties of the material. However, it is natural to ask about other cases where the muon has a strong effect on its surroundings.

\begin{figure}
    \centering
    \includegraphics[width=\columnwidth]{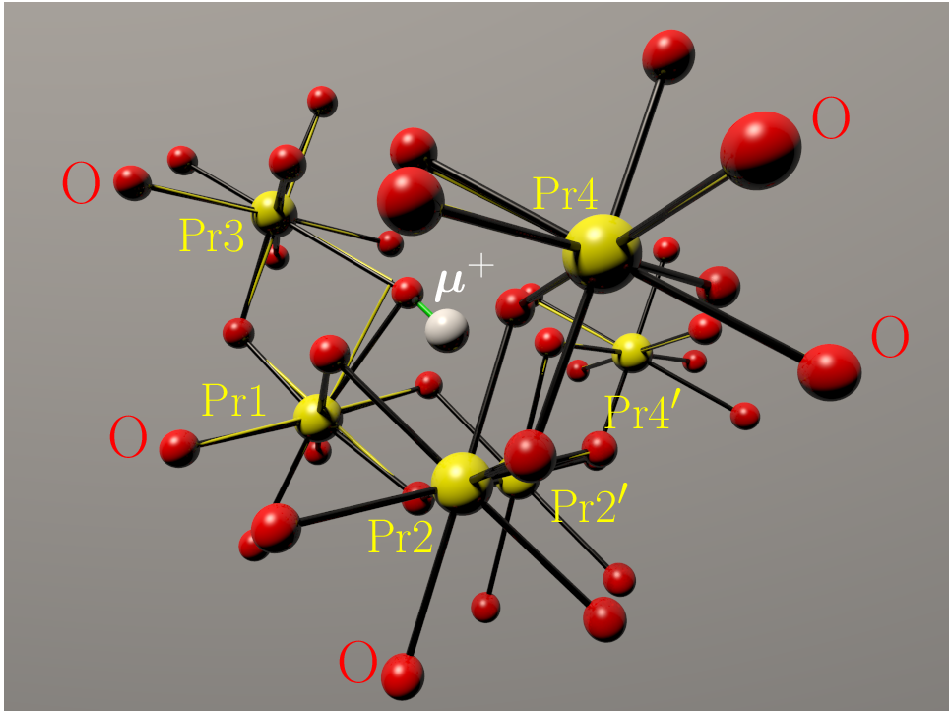}
    \caption{Muon stopping site (white sphere) and
atomic positions (oxygen ions are red, Pr ions are yellow, and Sn is suppressed) in the quantum spin ice Pr$_2$Sn$_2$O$_7$ as calculated by 
DFT$+\mu$. Bonds are shown as solid black lines, with the bonds of the unperturbed lattice (yellow lines)  included for comparison. Pr ions are numbered in order of separation from $\mu$ with Pr1 being the closest. The pair Pr2 and Pr 2' is equidistant from the muon, as are Pr4 and Pr 4'. [Adapted from F. R. Foronda {\sl et al.} Phys. Rev. B {\bf 92}, 134517 (2015).\cite{Foronda2015}] Copyright 2015 American Physical Society.}
    \label{fig:Pr}
\end{figure}

An important example of this was found in a family of quantum spin ices, a muon-induced change of the local crystal field dominates the measured response in a way that could be fully quantified,\cite{Foronda2015} demonstrating that the concerns raised above can be of critical importance.  Spin ice materials are based on the pyrochlore structure and the physics of such pyrochlore materials, with general formula A$_2$B$_2$O$_7$ is strongly dependent on the crystal field surrounding the lanthanide cation A.  Thus the introduction of a muon into the structure could alter that crystal field and change the nature of the ground state.

This effect is realised
in the material Pr$_2$Sn$_2$O$_7$, which exhibits quantum spin ice behaviour.\cite{Zhou2008} Without the muon, the Pr$^{3+}$ ion is surrounded by eight oxygen anions and adopts a doublet crystal field ground state, resulting in an effective spin-$\frac{1}{2}$ moment.\cite{Princep2013} However, DFT$+\mu$ calculations imply that the presence of the muon [the site is identified as $(-0.013, 0.047, 0.203)$] distorts the local symmetry around nearby Pr ions.  The largest effect is found to be an anisotropic distortion, with one Pr-–O bond strongly rotated (bent) and another significantly extended (see Fig.~\ref{fig:Pr}). The net result is that the doublet ground state is split on each of the neighbouring Pr cations has a singlet ground state. Then, via a hyperfine enhancement mechanism,\cite{Bleaney1973,blundell2023} a distribution of static magnetic moments is produced in which the average moment size grows on cooling. The model (involving DFT$+\mu$ calculations of the distortion and crystal field calculations of the resulting Pr environments) produced quantitative agreement with the temperature dependence of the observed muon response, demonstrating the validity of this approach.\cite{Foronda2015}  The Pr$^{3+}$ ion has a non-Kramers ground state and so is particularly vulnerable to electrostatic perturbations.  For the case of Dy$_2$Ti$_2$O$_7$, the Kramers doublet ground state of Dy$^{3+}$ is unaffected by the muon, even though the muon site and resulting anisotropic distortion are pretty much identical to that shown in Fig.~\ref{fig:Pr} for Pr$_2$Sn$_2$O$_7$.
\subsection{Quantum effects}\label{sec:quantumeffects}
The DFT treatments described so far
have assumed that the muon is a classical particle sitting statically at its
equilibrium position. This is a consequence of the Born-Oppenheimer approximation where we make a clear distinction between the classical nuclei (including the positive muon, which is treated as a light proton) and the quantum electrons. 
However, this approach neglects quantum
 mechanical effects of muons and nuclei, such as zero-point motion, tunnelling, and large zero-point energies (ZPEs).  These can change the energetic ordering of muon sites, destabilize certain  classical muon sites or cause several sites to merge into one if ZPE overcomes the potential-energy barriers between them. For nuceli, all of these effects are mass dependent and therefore especially pronounced for light particles such as the muon. 
 The result is that, in reality, the muon wavefunction is expected to be spread over an appreciable volume in the material, rather than concentrated at a single point.
 All observable quantities such as  muon coupling constants have, in principle, to be suitably averaged over the whole extended muon wavefunction, rather than sampled at a single classical point as in the Born-Oppenheimer approximation.
 The significant spatial extent of the
 wavefunction of the implanted muon is known to be relevant in many
 systems.  Well known examples are the quantum diffusion of muons in
 metallic systems such as copper or of muonium in insulating systems such as solid nitrogen. 
 
 Dealing with quantum effects is, of course, a difficult problem (DFT being a scheme invented to avoid this!) and so several approximations have been proposed and attempted. 
Perhaps the simplest class of approximation is a  harmonic one. This starts with a DFT
  calculation of the ($\Gamma$-point) phonon spectrum  for the classical muon site geometry.
An additional assumption often made is that, since muons are lighter than
nuclei,  muon zero-point motion is adiabatically decoupled from
the lattice over some relatively short length scale.

An example is, once again, provided by the fluorine states described in Section~\ref{sec:fluorides}. In vacuum the linear F--$\mu$--F anion has four vibrational modes: a symmetric stretch, bending
(two-fold degenerate), and an asymmetric stretch. In a solid the
twofold degeneracy of the bending mode is broken due to the
symmetry of the site: in LiF, NaF, CaF$_{2}$, and BaF$_{2}$ there are
two fundamentally inequivalent directions of bending; one is
towards a neighboring cation and is shifted up in frequency
while the other direction is into a gap in the crystal structure
and is shifted down in frequency. From the frequencies of these decoupled vibrational
modes we can estimate the zero-point energy (ZPE) of the
muon. The strong bond in this case,  combined with the small  muon mass
leads to a large ZPE for the F--$\mu$--F center
of 0.80~eV in vacuum. This is larger than the ZPE of any
natural triatomic molecule (the ZPEs of H$_{2}$O and H$^{+}_{3}$ are 0.56 and 0.54~eV, respectively),  demonstrating the importance
of quantum effects in muon localization.\cite{Moller2013}

More generally, the quantum muon problem can be addressed using a range of single-particle approximations where the muon sits in the potential of its surroundings and the Schr\"{o}dinger equation is solved for the muon only. These schemes can be most easily addressed in two limits: weakly- and strongly-bound muons. In both
of these, an effective single-particle potential
 is constructed from total DFT energy. To do this, the 
muon is displaced from its classical site, while: (i)
keeping the nuclei fixed at the positions corresponding
to the unperturbed muon site (in the weakly-bound case), or
(ii) letting the nuclei relax  to new lowest-energy
positions (in the strongly-bound case), while keeping
the center of mass of the system  fixed. 

\begin{figure}
\includegraphics[width=\columnwidth]{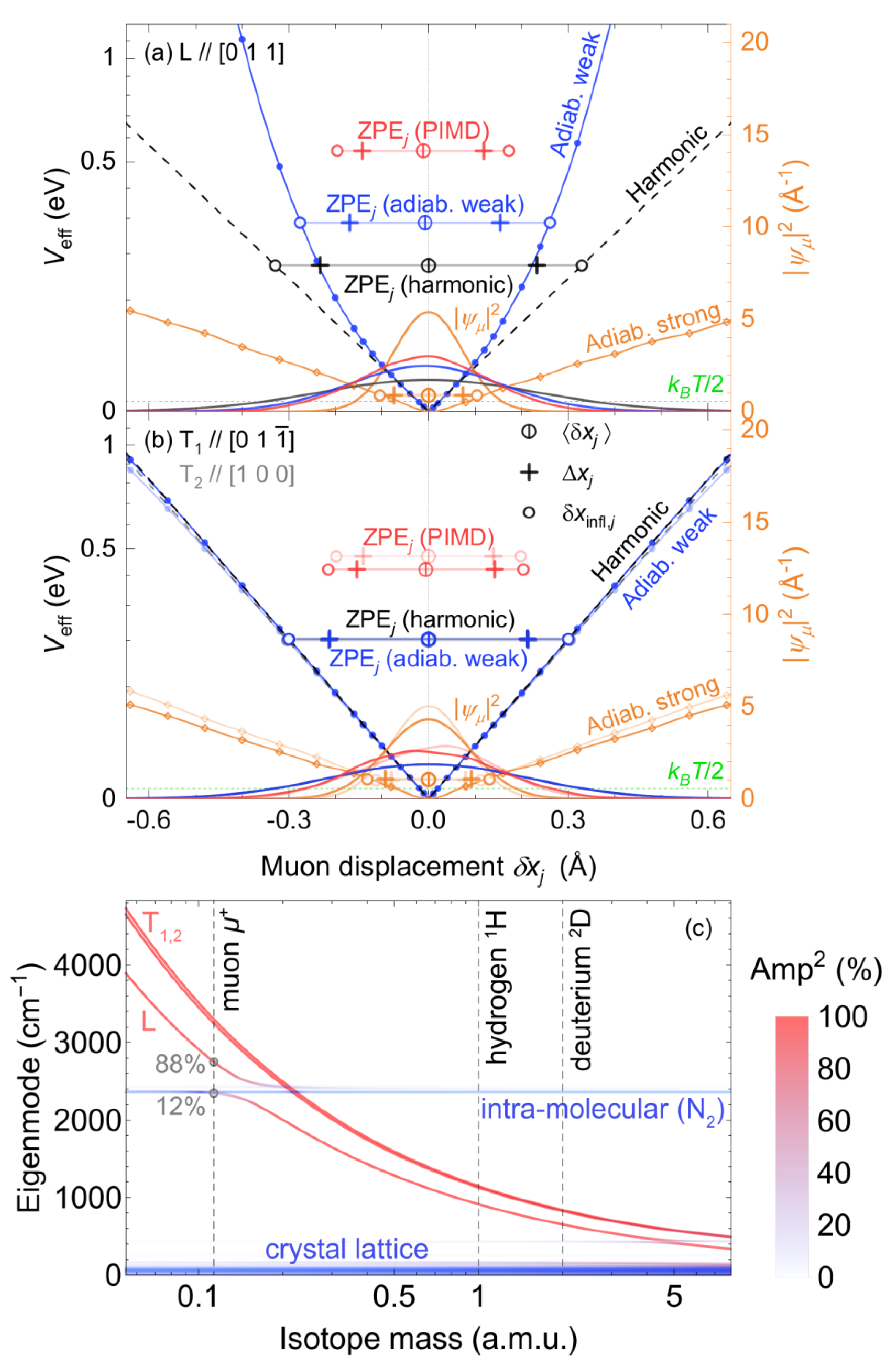}
\caption{\label{fig:n2} 
(a,b) Effective muon potential in $\alpha$--N$_{2}$ on a square-root linear scale (i.e.\ harmonic potentials appear as straight lines) along different directions for different approximations (left axis) with the corresponding muon probability density (right axis), directional contributions, mean muon displacements, wavefunction widths and inflection points. (c) $\Gamma$-point phonon spectra for muonated $\alpha$–N$_{2}$ vs.\ muon mass. Shading gives the muon amplitude squared in the normal modes. [Taken from M.~Gomil\v{s}ek {\it et al.}, arXiv:2202.05859.\cite{Gomilsek2022}]}
\end{figure}

An example of an attempt to apply these approaches is the case of $\alpha$-N$_{2}$.\cite{Gomilsek2022} Here the muon forms an extended electric-dipole polaron around a central, quantum-entangled [N$_{2}$--$\mu$--N$_{2}$]$^{+}$ complex.
The ZPE in the approximation schemes is shown in Fig.~\ref{fig:n2}(a) and (b).  The muon wavefunction is estimated to be significant on the length scale of the underlying crystal, making the quantum calculation necessary. Unfortunately, the two single-particle schemes give rather different estimates of the ZPE. The phonon calculation [shown in Fig.~\ref{fig:n2}(c) as a function of muon mass] shows the problem here: a strong hybridization of the longitudinal muon normal mode with intra-molecular vibrations of both N$_{2}$ in the [N$_{2}$--$\mu$--N$_{2}$]$^{+}$ complex, which implies significant muon–nuclear entanglement. As a result, both the single particle and harmonic approximations provide only a very rough guide to the quantum effects and do not agree between themselves. 
In cases such as that of N$_{2}$, an exact approach is therefore needed. This is carried out using path-integral molecular dynamics (PIMD), in which observables are calculated from arbitrary muon–nuclear zero-point motion. This scheme, based on discretizing an imaginary-time path integral, is very intensive on computer power, but can yield reliable numerical estimates. In the case of N$_{2}$ it confirms that the muon and its surroundings are strongly entangled, but leads to a very precise estimate of the $^{14}$N nuclear quadrupolar coupling constant.\cite{Gomilsek2022}

Continuing the theme of quantum mechanical effects, 
transition-state searches provide a route to evaluating the possibility of muon diffusion between candidate sites (an early example of this was applied to the study of muon diffusion in copper\cite{Bonfa2015}). 
In a general geometry optimization process,  the coordinates of the atoms are adjusted so that the energy of the structure is brought to a stationary point, in which the forces on the atoms are small. A {\it transition state} (TS) is a stationary point that is an energy maximum in one direction  and an energy minimum in all other directions.
A procedure that automates a search for TSs in codes such as {\sc Castep} \cite{clark2005} is particularly useful for predicting barriers to chemical reactions and determining reaction pathways.
In a chemical reaction, for example, starting from  reactants, energy increases to a maximum and then decreases to the energy of the products. The maximum energy along the reaction pathway is  the activation energy and the structure corresponding to this energy is  the transition state.
 These ideas can be used to find diffusion barriers, which is where its use in $\mu$SR arises.
 
\begin{figure}
\includegraphics[width=0.9\columnwidth]{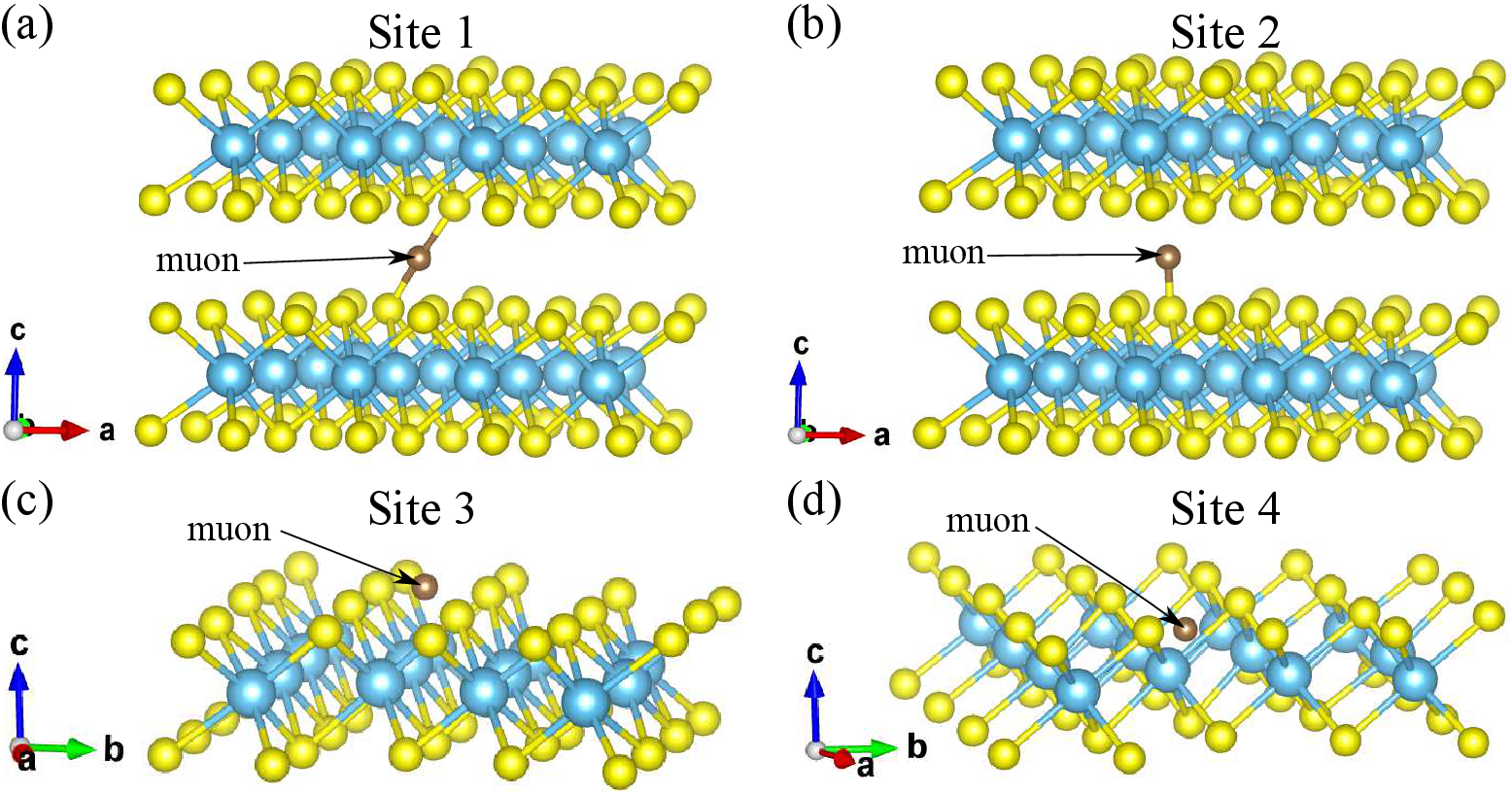}
\caption{\label{fig:tas2} Muon stopping sites in TaS$_{2}$. Ta and S are represented by blue and yellow spheres respectively. Sites 1 and 2 occupy the interlayer region. Site 3 is associated with the S layer and site 4 with the Ta layer. [Taken from F.~L.\ Pratt {\it et al.}, npj Quantum Materials {\bf 6}, 69 (2021).\cite{Pratt2021}]}
\end{figure}

A recent example of the use of 
transition-state analysis to evaluate muon diffusion in a solid 
is the case of TaS$_{2}$.\cite{Pratt2021} Here the muon
 sites form four distinct groups (Fig.~\ref{fig:tas2}). Muons in site 1  bond to one S atom in each of two adjacent TaS$_{2}$ layers forming a linear S--$\mu$--S state with unequal bond lengths. In site 2  the muon is bonded to only a single S atom (with a bond length of 1.4\,\AA), with these sites being around 0.03~eV higher in energy than site 1. In site 3, the muon stops inside the S layer, slightly displaced from the centre of the triangle defined by three S atoms. For site 4  the muon site is displaced by around 0.7\,\AA\ along the $c$-axis above the centre of the triangle defined by three Ta atoms.
To investigate the possibility of muon diffusion, TS searches between each pair of distinct muon stopping sites associated with a single TaS$_{2}$ layer were carried out. The energy barrier between sites 1 and 2 is very close to the difference in energy between the two. This, coupled with the large estimated  ZPEs  of both sites (0.35~eV and 0.51~eV respectively), means that the muon is likely to be delocalised between these two geometries, rather than there being two distinct stopping sites. Similarly, the barrier between sites 2 and 3 is less than their ZPE. Thus, sites 1, 2 and 3 are expected to form a single quantum delocalised state. 

\section{Outlook and conclusion}
DFT+$\mu$ is computationally costly, particularly for complex materials, and each new material investigated has to be studied from scratch in the same way. Many of the most exciting materials that are currently being discovered (including spin liquids) tend to be chemically complex, and this puts severe demands on the feasibility of electronic structure calculations.  Although we have concentrated on calculations of the muon site and associated local distortions, DFT calculations can also be used to estimate muon contact hyperfine fields in metals.\cite{Onuorah2018,Onuorah2019}

The $\mu$SR technique concentrates on using the positive muon, but experiments with negative muons can also be performed.\cite{blundell2022}  There is no problem with identifying the $\mu^-$ site because it is captured by the nucleus.  This does result in an atom with atomic number $Z$ behaving as a muonic atom with effective nuclear charge $+(Z-1)e$, the $\mu^-$ partially screening the full nuclear charge.  The effect on this state on the local crystal structure can then be calculated using DFT.\cite{Gill2023}  (This method can be termed DFT$+\mu^-$ to distinguish it from the usual DFT$+\mu^+$ technique.)   

A notable feature of the method used in (positive muon) DFT$+\mu$ of randomly-initializing muons in sites and then relaxing the structure, is that it is rather unlike the real-life situation of the muon finding its stopping site by impinging on the sample with an energy of  4~MeV and then shedding energy through interactions with the solid, before eventually coming to rest.\cite{blundell2022} This means that it is not possible, given an energy-ordered list of calculated muon sites, to tell which is realised, nor to reliably estimate the relative occupancy of several multiply occupied sites. Ultimately a capture cross section for each site is needed, whose value will reflect not only the energy, but also some details of the final stages of the muon stopping process.  This is likely to be an avenue of future research whose solution will allow further information to be squeezed from results obtained using the $\mu$SR technique.  Nevertheless, the progress obtained within the last few years using DFT$+\mu$ has already transformed the way in which muon data are both interpreted and understood.  Though this method is in its infancy, it has become an indispensable part of how muon experiments are now conducted.

\begin{acknowledgments}
We are grateful to the following students, postdocs, and colleagues, who have collaborated with us in our research in this area and made important contributions to this field:
Pietro Bonf\`{a},
Davide Ceresoli,
Stewart Clark,
Roberto De Renzi,
Marina Filip,
Francesca Foronda,
George Gill,
Matja\v{z} Gomil\v{s}ek,
Zachary Hawkhead,
Alberto Hernandez Mel\'{i}an,
Thomas Hicken,
Benjamin Huddart,
Dominik Jochym,
Mohammad Maikudi Isah,
Franziska Kirschner,
Franz Lang,
Leandro Liborio,
Nicola Marzari,
Johannes M\"oller,
Ifeanyi John Onuorah,
Francis Pratt,
Samuele Sanna,
Simone Sturniolo,
Johnny Wilkinson, and Hank Wu.   
We acknowledge funding from EPSRC(UK).  SJB acknowledges funding from UK Research and Innovation (UKRI) under the UK government’s Horizon Europe funding guarantee [grant number EP/X025861/1].
\end{acknowledgments}


\providecommand{\noopsort}[1]{}\providecommand{\singleletter}[1]{#1}%

\end{document}